\documentclass{aa}  

\usepackage{graphicx}
\usepackage{txfonts}
\usepackage{lipsum}
\usepackage{subcaption}         
\usepackage{lscape}             
\usepackage{placeins}           
                                
\newcommand{\kms}{\text{ km } \text{s}^{-1}}

\begin{document}

\title{What goes around comes around: \\the fate of stars in stripped tails of gas}

\titlerunning{Fate of tail stars}
\authorrunning{N. Akerman et al.}

\author{N. Akerman\inst{1,2}
    \and S. Tonnesen\inst{3}
    \and B. Poggianti\inst{1}
    \and R. Smith\inst{4}
    \and A. Werle\inst{1}
    \and E. Giunchi\inst{1,2,5}
    \and B. Vulcani\inst{3}
    \and J. Fritz\inst{6}
    }

\institute{INAF - Astronomical Observatory of Padova, vicolo dell'Osservatorio 5, IT-35122 Padova, Italy\\
         \email{nina.akerman@studenti.unipd.it}
        \and Dipartimento di Fisica e Astronomia `Galileo Galilei', Universit\`a di Padova, vicolo dell'Osservatorio 3, IT-35122, Padova, Italy
        \and Flatiron Institute, CCA, 162 5th Avenue, New York, NY 10010, USA
        \and Departamento de Física, Universidad Técnica Federico Santa María, Vicuña Mackenna 3939, San Joaquín, Santiago de Chile
        \and Dipartimento di Fisica e Astronomia `Augusto Righi', Universit\`a di Bologna, Via Piero Gobetti 93/2, 40129 Bologna, Italy
        \and Instituto de Radioastronomía y Astrofísica, UNAM, Campus Morelia, A.P. 3-72, C.P. 58089, Mexico
        }

\date{Received XX January 2025}

\abstract{
We conduct high-resolution wind-tunnel simulations to study in-situ star formation in the stripped tails of two massive ($M_\text{star}=10^{11}M_\odot$) galaxies undergoing time-evolving ram-pressure stripping: one is stripped face-on (W0) and the other is subject to an angled wind (W45). We find that the majority of stars in the tail are formed close to the galaxy disc at the beginning of stripping. Most stars have ages that reflect outside-in stripping -- older stars are found at larger radii than younger stars. The velocities and metallicities of stars indicate that ICM mixing both increases the velocity and decreases the metallicity of star-forming gas, leading to faster, lower metallicity stars at larger distances from the galaxies. However, not all stars follow this simple model, even in the case of face-on stripping. Indeed, a considerable number (15--25 per cent) of tail stars are formed with negative velocities, indicating the fallback of star-forming gas on to the galaxy. Almost all of the tail stars formed within 20 kpc from the disc will eventually fall back on to the galaxy, and their contribution to the intracluster light is negligible. The orbits of the stars formed in the tail result in an extended (and asymmetric in the case of W45) stellar distribution around the disc. Mock UV images reveal that the observed vertical distribution of its stars is not significantly broader than in an undisturbed galaxy, indicating that more stars would need to form in the stripped tail than we find in our simulations to observably impact the UV disc width of ram pressure stripped galaxies.}


\maketitle

\section{Introduction}

Properties of a galaxy (such as mass, morphology, metallicity, star formation, etc.) are influenced by the environment in which it resides \citep{Dressler1980,Poggianti1999,Goto03}. In dense environments such as galaxy clusters, a galaxy is subject to various processes, including ram pressure stripping \citep[RPS][]{GunnGott1972}. On its infall into a cluster, a galaxy is subject to ram pressure from the intracluster medium (ICM) which is able to remove the interstellar medium (ISM) of the galaxy in an outside-in fashion \citep[i.e. stripping first affects the galaxy outskirts before moving closer to the centre,][]{Quilis00, RoedigerBruggen06, Kronberger08, Cortese12, Cramer19, Owers19, Vulcani20a_GASP_XXIV}. The stripped material arranges into tails of gas that trail the galaxy and can be observed at different wavelengths: H$\alpha$ \citep{Gavazzi01, Smith10, Poggianti19_GASP_XXIII, Moretti22}, neutral \textsc{H i} \citep{Kenney04, OosterloovanGorkom05, Chung07, TonnesenBryan10, Poggianti19_GASP_XXIII}, UV \citep{Kenney14,George18}, X-ray \citep{SunVikhlinin05, Machacek05, Sun10, Tonnesen11}, in various bands with Hubble Space Telescope (HST) \citep{Owers12, Kenney15, Ebeling19, Durret21}, or as molecular gas \citep{Jachym17,Jachym19,Moretti18b_GASP_X}. While the gas removal means that star formation will inevitably quench in the disc \citep{Vollmer01, Gullieuszik17_GASP_IV, Owers19, Boselli20}, it can first induce star formation in both the surviving disc and the newly-stripped tail \citep{Vulcani18b, Vulcani20a_GASP_XXIV, Vulcani24}. 

Recent observations have revealed that the stripped tails do not merely dissipate into the ICM, but are able to form stars which are best seen in H$\alpha$ \citep{Sun07, Sun10, Smith10, Merluzzi13, Kenney14, Bellhouse17_GASP_II, Gullieuszik17_GASP_IV, Moretti18a_GASP_V, Moretti20b_GASP_XXII, Jachym19, Poggianti19_GASP_XXIII} and UV \citep{Fritz17_GASP_III, George18, George23}. The existence of in-situ tail star formation highlights the multiphase nature of tails, where there is enough dust and molecular gas to collapse and form stars embedded in the hot X-ray-emitting gas. Negative metallicity gradients along the tails \citep{Gullieuszik17_GASP_IV, Poggianti17_GASP_I, Bellhouse19_GASP_XV, Franchetto21a} suggest that two mechanisms are at play simultaneously: outside-in stripping (central regions of a galaxy have higher metallicity than the galaxy outskirts) and mixing of the stripped ISM with low-metallicity ICM \citep{Franchetto21b}. The stripped gas cools down and collapses to form stars at an efficiency that also decreases with distance from the galaxy disc \citep{Moretti20b_GASP_XXII}. These stars contribute from a few to 20 per cent of the total star formation in the disc + tail system \citep{Poggianti19_GASP_XXIII, Gullieuszik20_GASP_XXI, Werle22}. In-situ tail star formation has also been studied in simulations, albeit without a consensus. While some do find that the stripped gas can form stars \citep{Kapferer09, TonnesenBryan12, Roediger14, Lee22}, others do not \citep{Bekki14,Steinhauser16,Lee20}. 

Star-forming clumps have also been observed with unprecedented resolution using HST allowing for a statistical study of star formation in tails and deepening our understanding of star formation processes \citep{AbramsonKenney14, Cramer19, Giunchi23a, Gullieuszik23}. \cite{Giunchi23b} present systematic evidence for the fireball model, in which a clump continues to actively form stars as it is being stripped so that this stream of stars forms a gradient of stellar ages \citep{Cortese07, Kenney14, Jachym17}. \cite{Werle24} look at mass, age and spatial distribution of star-forming clumps and find that clumps farther away from the galaxy are younger and less massive. The authors conclude that the stars form from molecular gas that was both directly stripped from the disc (close to the disc) and formed in-situ as a result of cooling of the stripped atomic gas (far away from the galaxy).

In this work, in order to compare with new observational discoveries, we make use of the high resolution of our simulations and study where and how stars form in the stripped tails and their fate. After briefly outlining the simulation suite in Sect. \ref{sec5:suite}, we look at star-formation histories and subsequent fallback of stars in Sect. \ref{sec5:SFH_fate}. In Sect. \ref{sec5:star_formation_how} we analyse where in the stripped tails stars are formed and how their metallicities can inform us about the star-forming gas and in Sect. \ref{sec5:uv_maps}, we present mock observational UV maps that can be compared to images taken with HST. Finally, in Sect. \ref{sec5:discussion} we discuss the results and their implications for observational studies and summarise our conclusions in Sect. \ref{sec5:conclusions}.

\section{Suite of simulations} \label{sec5:suite}

We introduce the simulations in detail in our previous work, \cite{Akerman23}, while here we summarise the main features. Our simulations employ the adaptive mesh refinement code \textsc{Enzo} \citep{Enzo}. The simulation box has 160 kpc on a side and includes 5 levels of refinement allowing for a maximum resolution of 39 pc. The refinement criteria are Jeans length and a baryon mass of $\approx7500 M_\odot$.

The simulations include radiative cooling using the {\sc grackle} library \citep{Grackle} with metal cooling and the UV background by Haardt \& Madau \citep{HaardtMadau12}. The ISM starts with a metallicity $Z = 1.0 \; Z_\odot$ \citep[observed for galaxies of this mass,][]{Mannucci10} and the ICM has $Z = 0.3 \; Z_\odot$.

The star formation and stellar feedback recipes are implemented as in \cite{Goldbaum15, Goldbaum16}. If the mass of a cell exceeds the Jeans mass and the minimum threshold number density of $n_{\rm{thresh}}=10 \, \rm{cm}^{-3}$, a stellar particle will form with a minimum mass of $10^3 M_\odot$, assuming a star formation efficiency of 1 per cent. The feedback includes momentum and energy input from supernovae (SNe), ionising radiation from young stars (heating up to $10^4$ K) and winds from evolved massive stars. SNe have combined energy budget of $10^{51}$ erg which is first added as the terminal momentum input to the 26 nearest neighbour cells and any additional energy is deposited as thermal energy to the cell containing the SN. SNe also increase the metallicity in the host cell.

To model a galaxy, while the self-gravity of the gas component is calculated at each time step, we use static potentials for the stellar disc and the spherical dark matter halo \citep[see][]{RoedigerBruggen06, TonnesenBryan09}. The model galaxy is based on a well-studied RPS galaxy, JO201 \citep{Bellhouse17_GASP_II, Bellhouse19_GASP_XV}, which is only used to select realistic initial conditions. For the Plummer–Kuzmin stellar disc \citep{MiyamotoNagai1975} we use a mass of $M_\text{star} = 10^{11} M_\odot$, a scale length of $r_\text{star} = 5.94$ kpc and a scale height of $z_\text{star} = 0.58$ kpc. We model a Burkert profile for the dark matter halo \citep{Burkert1995, MoriBurkert00} with a core radius of $r_\text{DM} = 17.36$ kpc. Initial parameters of the gaseous disc are the following: mass $M_\text{gas} = 10^{10} M_\odot$, scale length $r_\text{gas} = 10.1$ kpc and scale height $z_\text{gas} = 0.97$ kpc.

To simulate RPS, the galaxy is fixed in the centre of the simulation box and an ICM wind is added via inflow boundary conditions (and outflow on the opposite side). We include a time-varying (in density and velocity) ICM wind \citep{Tonnesen19}, and to find its parameters, we model a galaxy on its infall into a massive cluster ($M_\text{cluster} = 10^{15} M_\odot$) following the procedure described in \cite{Bellhouse19_GASP_XV} from a clustercentric radius of 1.9 Mpc and with an initial velocity of $1785 \; \mathrm {km \, s}^{-1}$. The ICM has a beta profile, is isothermal ($T=7.55\times10^7$ K) and is in hydrostatic equilibrium. From Rankine–Hugoniot jump conditions for Mach number of 3 we find the pre-wind ICM conditions using the ICM wind parameters at the initial radius. Notice that this set‐up does not include a circumgalactic medium surrounding our modelled satellite galaxy \citep[although 0.1 per cent of the original ISM can reach 5 kpc above the disc due to feedback,][]{Akerman24}, and the galaxy transitions straight into the ICM.

Since the galaxy is fixed in space, we define a wind angle as the angle between the wind direction and the galaxy rotation axis, and model two wind angles: $0^\circ$ (a face-on wind that flows along the $z$-axis in the simulated box, W0), $45^\circ$ (W45, in which the wind has equal components along the $y$- and $z$-axes). The wind angle is constant throughout a simulation. As a control, we also simulate a galaxy that does not undergo RPS (no wind, NW).

We evolve our galaxies in isolation for 230 Myr in order to stabilise the SFR, such that the variation of the SFR on a 5 Myr time-scale decreases to 5 per cent. The wind reaches the galaxies after 70 Myr, therefore the galaxies evolve in total for 300 Myr before the onset of RP.

Finally, throughout this work, we define the tail region using $z = [2, 20]$ kpc. The lower boundary is based on the galaxy disc definition from previous work \citep{Akerman24} where the disc height was set to $z=\pm2$ kpc. The upper boundary is motivated by refinement restrictions, according to which the gas below 20 kpc is allowed to be resolved up to 39 pc, while the gas above 20 kpc is always on the lowest refinement level with a resolution of 1248 pc.

\section{Star formation history and the fate of stars in the tail region} \label{sec5:SFH_fate}

\begin{figure}
\centering
\includegraphics[width=\hsize]{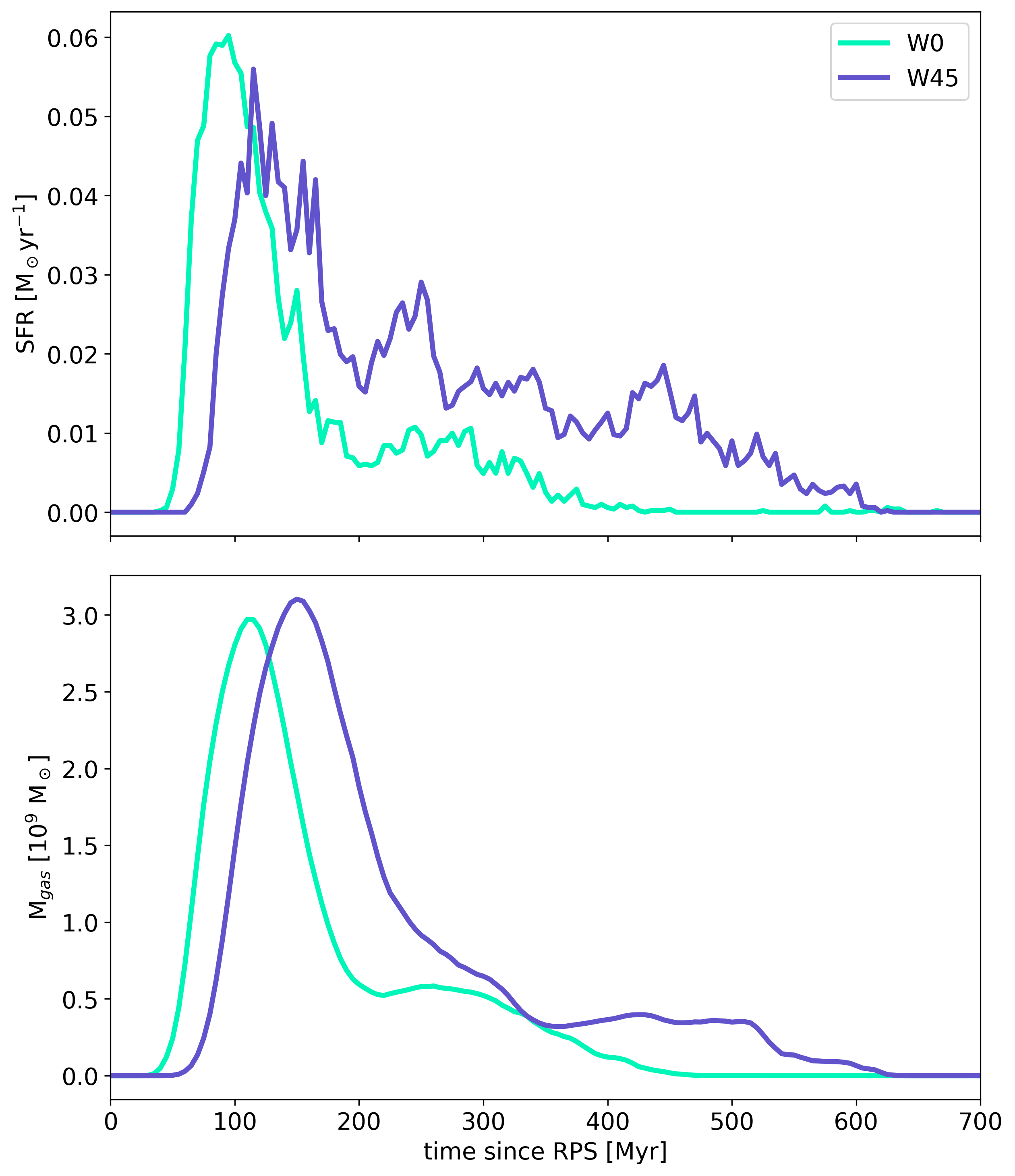}
\caption{\textit{Top}: star formation history, \textit{bottom}: mass of cold ($T<10^{4.5}$K) gas as a function of time, in the tail region ($z = [2, 20]$ kpc), colour-coded by the wind angle.}
\label{fig:SFH_tail}
\end{figure}

In this section we determine how many stars form in the stripped tails of our massive galaxies, where they form within the tail and how their positions evolve over time. We will first present the global star formation rates in the W0 and W45 tails. We then show the distribution of where stars are formed in the tail region, and where they `end up' once the galaxies are completely stripped of their ISM 700 Myr after RPS has begun.  

Let us first examine the star formation history in the tail region shown in the top panel of Fig. \ref{fig:SFH_tail}. By the end of the simulation a total of 5871 and 9270 tail star particles form in W0 and W45, respectively, which equate to stellar masses of $\sim6\times10^6 M_\odot$ and $\sim10^7 M_\odot$ (keep in mind that each particle has a mass of $10^3 M_\odot$). The larger number of star formed in W45 compared to W0 corresponds to a larger fraction of star formation in the tail, with tail stars constituting 0.88 and 1.02 per cent of all stars formed after the onset of RPS for W0 and W45, respectively. The two star formation histories closely follow their respective stripping rates discussed in detail in \cite{Akerman24}, and, by extension, the mass of the cold gas ($T<10^{4.5}$K) in the tail region (bottom panel). Note the 25 Myr delay between W45 and W0 which is simply due to the fact that it takes the angled wind more time to reach the galaxy since it has a larger distance to travel to the centre of the simulation box. What is important here is that while the stripping rates of the two galaxies are quite similar, W45 surprisingly has a higher tail SFR during most of the simulation (though the peak star formation is higher in W0) and for a prolonged period of time (even accounting for the aforementioned delay), while W0 practically quenches tail star formation after 400 Myr. To a lesser degree, this behaviour is also reflected in the cold gas mass.

\begin{figure}
\centering
\includegraphics[width=\hsize]{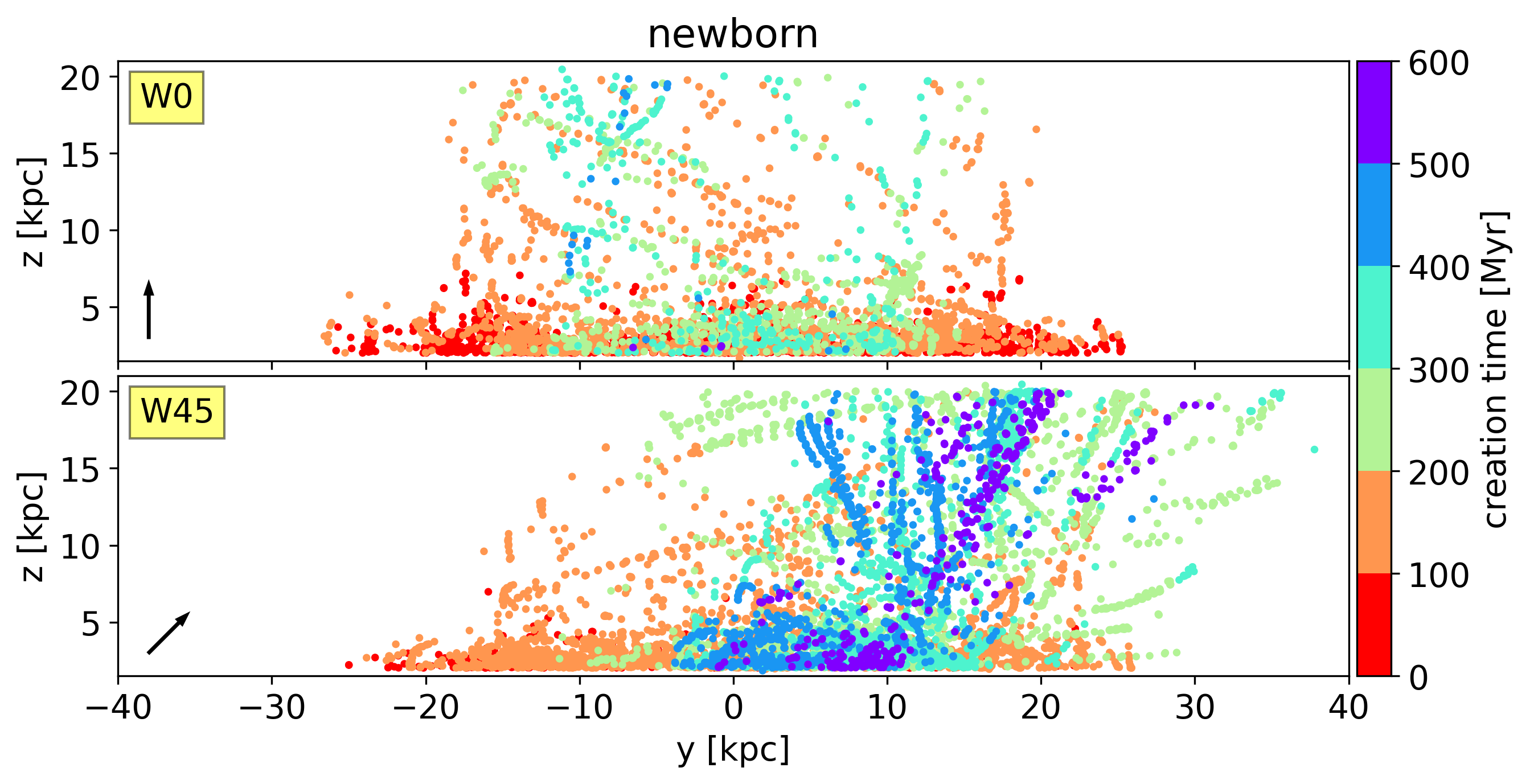}
\caption{For W0 (top) and W45 (bottom): position of stellar particles at the moment of their birth, colour-coded by the creation time (since RPS). Arrows indicate wind direction. Star formation proceeds outside-in for both wind directions, and shows significantly more symmetry in W0 than in W45.}
\label{fig:tail_birth}
\end{figure}

To understand where in the wake of the galaxies star formation happens and what drives the differences between the two galaxies, in Fig. \ref{fig:tail_birth} we plot the position of stars at the moment of their birth for W0 (top) and W45 (bottom). The points are colour-coded by the time (since RPS) at which stars were born. We emphasise that this distribution of stars is not a snapshot at any particular moment of time, but every star that is formed 2 kpc or more above the disc is plotted at the $[y, z]$ position at the time of its creation and colour-coded by the creation time.

\begin{figure}
\centering
\includegraphics[width=\hsize]{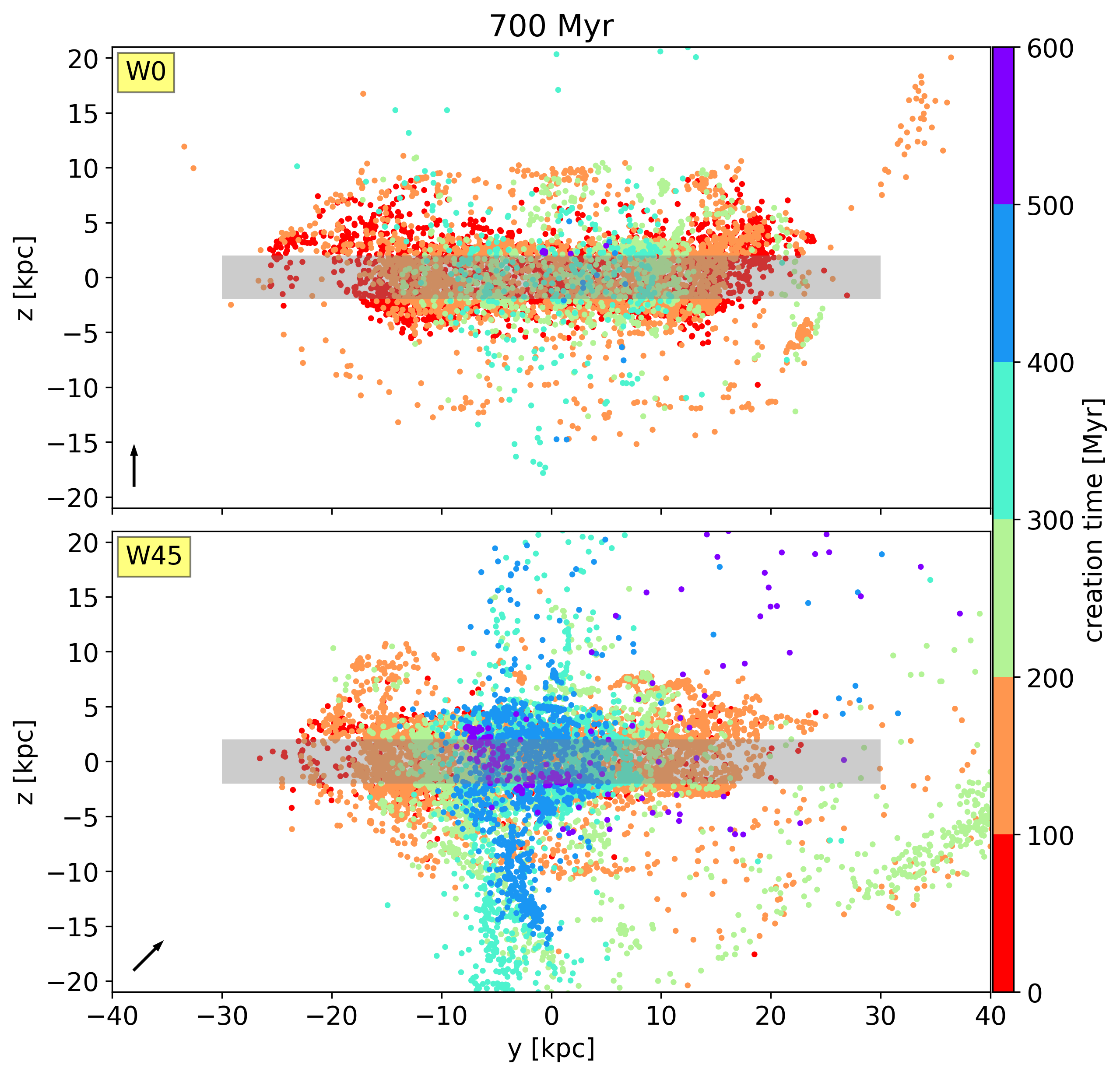}
\caption{For W0 (top) and W45 (bottom): the positions of stars at 700 Myr, colour-coded by the creation time (since RPS). The colours for the same points are identical to Fig. \ref{fig:tail_birth}. The grey area outlines the disc ($R=30$ kpc, $z=\pm2$ kpc), and arrows indicate wind direction.}
\label{fig:tail_evolve}
\end{figure}

Stripping in W0 proceeds symmetrically outside-in, which is also reflected in how stars are initially formed at large $y$-values (red points), and with time they start gradually being born within smaller and smaller cylindrical radii. It is also clear that the majority (80 per cent) of stars are born close to the disc ($z<5$ kpc). Although it is shown with a minority of stars, we also notice that the vertical distribution of stars changes with time. In the first 100 Myr (red points) all of the stars were formed within 8 kpc of the plane of the disc, as most of the stripped material has not yet reached large distances above the disc. Later in the simulation the gas tails stretch all the way up to (and beyond) 20 kpc, which is also reflected in the positions of newly-formed stars. This panel also illustrates that in W0 most tail star formation stops after 400 Myr, with only 29 star particles (or 0.5 per cent of the total) formed after that.

Looking closely at the stellar distribution above $\sim$5 kpc, one could also note that some of the stars seem to form connected sequences. Interestingly, most of these sequences seem to be of the same colour, indicating that the star formation along a path is of limited duration ($\le$100 Myr). We do note that a few sequences of two colours could indicate longer star formation episodes, although we only see one sequence spanning three time bins (which means the star formation must have lasted for more than 100 Myr). Most of these visually connected newly-formed stars seem to form diagonal sequences, even though the wind is flowing along the $z$-axis. We conclude that this is a reflection of the galaxy rotation that some of the gas still retains after being stripped. At larger radii we note that more of the visual stellar streams are vertical, which is a projection effect. While tracking individual star-forming clouds is beyond the scope of this work, we will discuss the gas- star connection in more detail in Sect. \ref{sec5:star_formation_how}.

Although (as already mentioned) the stripping rates themselves are quite similar between the two galaxies, in W45 stripping presents a more complex morphological picture. Here, the vertical distribution of stars is slightly more uniform compared to that of the face-on stripped galaxy, with 57 per cent of stars born within 5 kpc of the disc, and while we do see signatures of outside-in star formation as a result of gas removal, it is more clear to the left of the panel. In this figure, the ICM wind has components along the $z$- and $y$-axes, flowing in the direction of positive values. Once the gas is stripped, its movements are dictated by the combination of the ICM wind pushing upwards and along the $y$-axis and the retained rotation around the galaxy centre. This way, the gas moving along the wind direction quickly gets pushed to higher $y$-values (very extended stellar streams moving from negative to positive $y$). There, trying to continue in its orbital path, the gas tries to turn and move in the negative $y$ direction (against the wind), but is instead slowed down by the incoming ram pressure. As a result, dense gas is slowed down and spends a longer time in the region of high $y$-values (at the turning point of its orbit) where it forms stars. In projection, the resulting configuration of newly-formed stars is well illustrated by the blue and purple points in bottom panel of Fig. \ref{fig:tail_birth} (note that at the same time the wind also pushes the gas upwards).

We create a new projection to help us to understand what happens to the stars after they are formed. In Fig. \ref{fig:tail_evolve} we plot the final distribution of star particles as it appears at the end of the simulation at 700 Myr. Unlike Fig. \ref{fig:tail_birth}, this is a snapshot showing the position of all star particles that were formed above the disc (stars formed in the disc are not present here). As previously mentioned, W45 has longer-lasting and stronger tail star formation, particularly after 200 Myr, and these differences are reflected in the final distribution -- while stars formed in the first 200 Myr (red and orange) form relatively similar thick discs (more on that below), stars formed afterwards have completely different distributions, with those in W45 forming a structure akin to a `tail' under the disc, and those in W0 distributed more uniformly.

In both W0 and W45, under the influence of gravity most star particles fall back on to their galaxies, forming a thick disc with $z = \pm8$ kpc height. Some stars fall through the disc much further, even below -15 kpc in case of W45, which indicates that they formed from gas that was high above the disc and/or moving rapidly away from it and now have an orbit that is strongly misaligned with the original disc. Note that this image is but one snapshot in time -- left on their own, the particles both above and below the disc will continue to oscillate around it for a prolonged period of time. In Sect. \ref{sec5:uv_maps}, we will test whether this thickening can be observed with our current telescopes. Real stars born in-situ in the stripped tails are subject to the gravitational attraction not only from the galaxy but also the cluster. The wind-tunnel simulations do not account for the cluster's gravitational field, therefore almost all star particles fall back on to the galaxy\footnote{Only 3 star particles are able to achieve high enough velocities to escape the galaxy in W45, and none in W0. Thus, we can assume that this process would not contribute to a significant loss of tail-formed stars in real galaxies either.}. In Sect. \ref{subsec:ICL}, we consider the tidal stripping of stars in post-processing.

Looking at the upper panel of Fig. \ref{fig:tail_evolve}, one might think that after their formation the stellar particles fall `straight down' on the galaxy disc as in this projection the age segregation is still visible with the oldest stars (red points with the earliest creation time) located in the galaxy outskirts while the younger ones end up closer to the centre (keep in mind the projection effects). This is true for the majority (80 per cent) of the stars, while the others end up displaced from the radius of their formation to lower or higher radii. We will discuss this occurrence and the underlying mechanism in detail in Sect. \ref{subsec:radial}. Here, we emphasise that after fallback the stars do not concentrate in the galaxy centre.

The picture is similar in W45 (bottom panel of Fig. \ref{fig:tail_evolve}), but instead of falling straight down, the stars (due to the additional increase in the $y$-velocity component) follow a parabolic orbit which results in a more asymmetric distribution. The green points to the right of the panel follow a particularly elongated orbit, seemingly on their way to the galaxy centre. At 700 Myr there is an excess of stars below the disc, but in a longer simulation the stars would continue their orbit oscillating between being above and below the disc. Nevertheless, despite the complex stellar dynamics, the age segregation can also be seen even in the W45 galaxy.

\cite{Kapferer09} find that stars formed in the tails would fall on the galaxy centre forming a bulge, while \cite{TonnesenBryan12} measure the bulge-to-total stellar mass ratio and conclude that there is very little growth of the existing bulge in their galaxy undergoing RPS. Both works model a face-on stripped galaxy like W0. In our set-up we do not include a stellar bulge in the form of a static potential. Here, we find that after the fallback there is no significant mass concentration in the galaxy centre, especially in W0. In W45, stars that were formed later in the simulation (blue points) do end up in the centre of the disc. However, they do not form a visibly thicker area that could be identified as a bulge and the added mass measured in the sphere of $R=5$ kpc is only $3\times10^6 M_\odot$, insignificant compared to the total mass of the stellar disc ($10^{11} M_\odot$). We thus conclude that the tail stars are unlikely to form or contribute to a bulge, and drive a galaxy's morphological evolution from a spiral to an S0.

To summarise, both W0 and W45 galaxies are largely stripped outside-in, and the stars that form as a result of this process fall back on the galaxy almost `straight-on' maintaining the age gradient within the disc. They also seemingly form a thick disc, although it is uniform along the radius, with no major contribution to the bulge. More stars are formed when an angled wind is present as the gas gets slowed down and spends more time in one sector of the galaxy.

\section{When and how the stars are formed} \label{sec5:star_formation_how}

\begin{figure*}
\centering
\includegraphics[width=\textwidth]{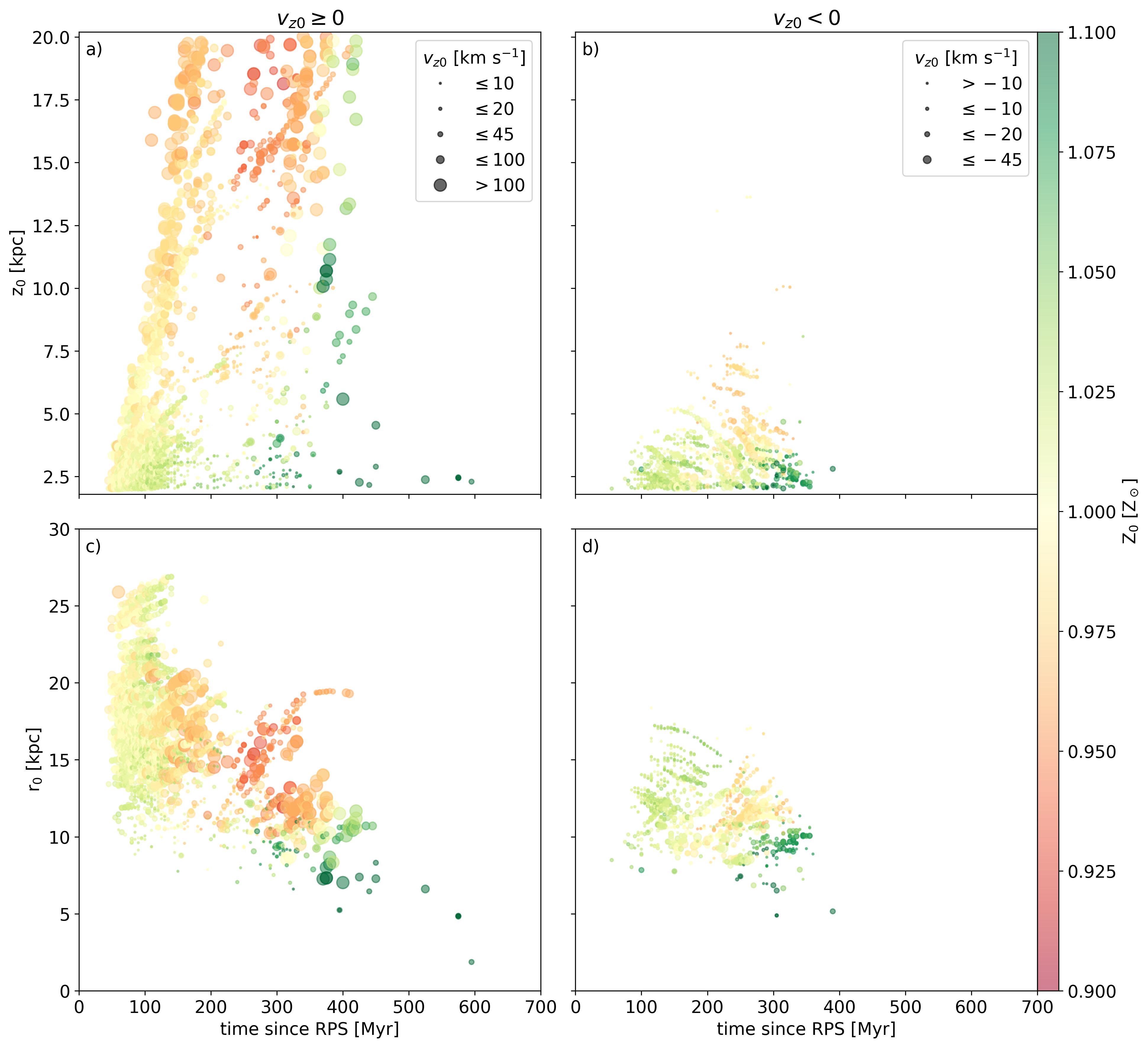}
\caption{For W0: vertical (top) and cylindrical r (bottom) distribution of newborn tail star particles as a function of time colour-coded by their metallicity. Left and right columns divide the stars into those born with positive and negative $v_{z0}$, respectively. Size of each point corresponds to $v_{z0}$ (in the legend).}
\label{fig:r_z_metallicity}
\end{figure*}

Now that we know the fate of stars formed in the inner 20 kpc of the stripped tail, we would like to go back to the moment of their birth and analyse the properties of gas that formed the stars, using the fact that stars adopt the properties of the gas from which they form. To do this, we will be looking at the stellar metallicities of the newly-formed particles and we will focus on the W0 galaxy as a more straightforward case. To keep our analysis more straightforward, we choose to exclude W45 from the discussion which introduces additional complexity in that it is stripped along both $z$- and $y$-axes. This factor makes the cylindrical radial position of star formation less clearly related to the stripping radius and various metallicity trends more difficult to disentangle, thus complicating drawing meaningful conclusions. In these simulations, metallicity of a newly-formed stellar particle is assigned based on the metallicity of the gas cell in which it was formed. Since in adaptive mesh refinement simulations following specific gas clumps can only be done using error-prone post-processing techniques, metallicity can be a useful property to study the mixing of ISM and ICM. Moreover, metallicity analysis is also used in observational studies of RPS \citep{Franchetto20, Franchetto21a, Franchetto21b}, and this will allow our findings to be directly applied to observational data.

For W0, Fig. \ref{fig:r_z_metallicity} shows vertical (top row) and radial (bottom row) distributions of newly-formed tail star particles as a function of time colour-coded by their metallicity. The columns divide the stars into those whose initial $v_{z0}$ is positive (with respect to the galaxy, left column) and negative (right column). The size of each point corresponds to $|v_{z0}|$, with bigger points indicating bigger absolute velocities (as described in the legend). The properties of each star particle correspond to the moment it was formed, i.e. in this figure, each star appears only once. In these simulations, the ISM starts with a metallicity of $1.0 Z_\odot$ and the ICM has $0.3 Z_\odot$ metallicity, so metallicities below the initial ISM value (red and orange colours) suggest that the gas that formed the stellar particles (`parent' gas) underwent some degree of mixing with ICM. In practice, since the ICM wind hits the galaxies after they underwent 300 Myr of evolution in isolation, the ISM metallicity is higher than $1.0 Z_\odot$ when the stripping begins (Appendix \ref{appendix:metallicity}). While this means that the gas with $Z>1.0 Z_\odot$ could still be partially mixed, later in the section we will discuss why in this figure we could still assume such gas to be `pure' ISM.

In this image, we can see the following trends:
\begin{enumerate}

\item Metallicity decreases with distance from disc plane as the stripped gas gets mixed-in with ICM. In panel a) we can notice two metallicity (colour) gradients: vertical and horizontal. Vertically (i.e. at a single time), stars formed farther from the disc get more metal-poor, suggesting that the more time the gas spends outside of the galaxy disc and the farther it reaches above it, the more it will mix with the low-metallicity ICM. Almost all stars formed above 5 kpc have metallicities below $1.0 Z_\odot$. The mixing is also evident from the velocities increasing with $z_0$ as this process imparts more momentum to the stripped gas \citep{TonnesenBryan21}. Conversely, horizontally, over time, metallicity of the stars increases suggesting that they are formed from gas that is stripped from progressively more metal-rich ISM (the low-metallicity points that seem to disrupt the trend will be discussed later).

\item Stars are unevenly distributed along the $z$-axis. In panel a) at early times, stars do not form far from the disc. Later, although at each point of time there are stars born at different heights, they are frequently unevenly distributed along the $z$-axis, with gaps such as the one between 7 and 12 kpc around 200 Myr\footnote{Please note that the bigger, faster points visually take up more space and might contribute to the bias.}.

\item Tail star formation avoids small radii ($r_0<10$ kpc) at early times because the galaxy is stripped from the outside-in. We also see in panel c) that only during the first 100 Myr do stars form at radii $r > 20$ kpc. While we can see the general trend of star particles forming closer to the galaxy centre as time progresses, which is indicative of outside-in stripping, some outliers (notably the stars formed from very mixed-in gas, orange and red points) can further complicate this picture. 
 
\item Stars do not form within the central few kpc of the rotation axis. In the same panel c) although the W0 galaxy has all of its gas removed, stars do not form $\ge$2 kpc above the disc plane within 5 kpc of the rotation axis (except for a single star at late times). Given that we know from \citet{Akerman23} that dense gas is in the central disc at late times we are left with two likely possibilities: either the dense gas is stripped into the tail, but mixing with the ICM causes it to heat and grow less dense so it cannot form stars, or the dense gas forms stars before it reaches 2 kpc above the disc. Future work examining these possibilities in detail by tracking dense gas would help distinguish between these scenarios. Importantly, this highlights that removing high-density gas from a galaxy via RPS does not necessarily result in star formation in the tail.

\item The initial burst of star formation is due to the initial `wave' of ram pressure passing through the disc. The most prominent feature in panel a) is the continuous `trail' of stars that starts at 50 Myr close to the disc and ends all the way up 20 kpc\footnote{This is the farthest point at which stars can form in our simulation due to the resolution requirement for star formation, see Sect. \ref{sec5:suite}} at 200 Myr. Moreover, these stars have quite high velocities (up to $100 \kms$) even though the stripping has just begun and panel c) reveals that these stars were formed with a wide range of $r$, which means that the gas was stripped not only from the outskirts but as close to the galaxy centre as $r = 10$ kpc. This time corresponds to the peak of SFR (Fig. \ref{fig:SFH_tail}), and from 0 to 150 Myr 70 per cent of all tail stars were formed. All these factors combined as well as the high stripping rate \citep[Fig.\ref{fig:SFH_tail}, bottom panel and][]{Akerman24} suggest that the burst of star formation is the result of the initial shock passing through the galaxy when the disc was hit with the wind front. The displaced gas continued to be gradually stripped away, mixing with the ICM in the process. 

\item A significant population of stars is formed as a result of gas fallback on to the galaxy from the disc shadow. In panel b) almost all of the stars with negative $v_{z0}$ are located close to the disc. 23 per cent of the stars are born with $v_{z0} < 0$, indicating significant fallback of stripped dense gas that survives to form stars. Indeed, in panels a) and b) it is possible through careful examination to identify several sequences of stars that form arches in $z<5$ kpc. We posit that gas was forming stars as it was stripped, reached the peak at $z=5$ kpc (panel a) and started falling on the galaxy, while continuing to form stars on its way back. In panel b) with time, the points in these arches also increase in size, as the gas is accelerated by the galaxy's gravity.  

Although gas stripping generally happens outside-in along the galaxy disc, stripping is possible within the theoretically defined stripping radius as the commonly-used definition by \cite{GunnGott1972} does not take into account the uneven distribution of surface densities within the disc itself and the existence of gas with surface densities as low as $10^{-4} \text{g cm}^{-2}$ even within the inner 5 kpc \citep[Fig. 4 from][]{Akerman24}. Thus, somewhat lower-density gas could be stripped from the disc, then move into the shadow of surviving denser gas and begin to fall back. If it then cools into a smaller, denser cloud the fallback could continue even if it is pushed again by the ICM wind. This picture is corroborated in panel d), as the fallback stars are formed at smaller $r_0$ than the stars with positive $v_{z0}$ (panel c), suggesting that their parent gas could indeed be protected from the ICM wind by the galaxy disc.

\item Stripped gas tends to flow in the outer radial direction due to the loss of centripetal force. If the gas survives long enough and is able to form stars along the way, they will have progressively increasing $r$ as well as increasing $z$, as illustrated by the very-low-metallicity stars (red and orange, 200--400 Myr) in panel c). 

This process can be reversed when the gas starts falling back on the galaxy as it experiences increase in centripetal force (light green points, panel d).

\end{enumerate}

Although we do not show a repeat figure in this section for W45, we note that all of the above main results are qualitatively extremely similar for the W45 run. Many more stars are formed, so while the `gaps' described in point 2 exist they are much smaller. In addition, while only 6 per cent of stars in W0 form in the inner part of the disc ($r<10$ kpc), in W45 galaxy this fraction increases to 12 per cent since angled ram pressure changes the orbits of the stripped gas. Because of this, radial evolution of the gas is also stronger, with some `trails' of stars being 10 to 25 kpc long in the radial direction, much more extended than the radial trails in Fig. \ref{fig:r_z_metallicity}, panel c). Finally, the fraction of stars formed as a result of gas fallback is somewhat lower, 15 per cent, although the absolute numbers of such stars between W0 and W45 are quite similar ($\sim 1370$ stars).

In this picture we are left with a few groups of star particles whose behaviour cannot be explained by the simple principles outlined above. For example, in panel d) there is a group of low-metallicity stars (orange) that despite falling back on the galaxy continue to move in the outward radial direction. One possible explanation lies in the fact that these stars are formed from well-mixed gas. Although in W0 the ICM wind has only the $z$-velocity component, upon hitting the galaxy, the wind front deviates in its route trying to go `around' the disc. This way, the wind gets additional (positive) cylindrical $r$-velocity components, that it can then impart on the stripped gas (this could also apply to the red points in panel c). Lastly in the same panel d), the origin of the very-high-metallicity (dark green) points is not clear either, as despite being formed outside of or very close to the stripping radius (which is around 8 kpc at 280--360 Myr), these star particles indicate that their parent gas was falling back on the galaxy (at low velocity, too) instead of being stripped away. We suspect that this signifies stars that formed from dense clouds that were nearly able to maintain their position in the disc. Because they were dense and small, their ICM fraction remained low, and they were able to continue radiatively cooling, thus increasing their density and reducing their radius. This reduction in the mixing surface resulted in less acceleration from the ICM wind and allowed for the clouds to slowly fallback, even while outside the galaxy's shadow. Until we can track a large population of individual clouds, perhaps in future work, we merely state this as a possible explanation for star formation from high-metallicity gas that is falling back towards the disc. These star particles along with the fallback stars and those that flare out due to the loss of centripetal force once again highlight that even in the case of face-on stripping, the dynamics of gas is not guided by simple outside-in stripping.

Finally, we address the fact that the raw metallicities might not be telling us the full story. While the galaxy evolves and forms stars in the disc, they will explode as SNe, which increases the ISM metallicity. This means that the disc gas, while starting with $1.0 Z_\odot$, will slowly become more and more metal-rich. In Appendix \ref{appendix:metallicity} we show this evolution in radial rings across the galaxy disc. As detailed there, the outer regions of the galaxy ($r>10$ kpc) the metallicity increases very slowly once RPS is introduced and stays around 1.0--1.05$Z_\odot$. Since most of the stars in the tail are formed outside of this radius, we find our assumption that metallicities $Z>1 Z_\odot$ are an indication of the `pure' or unmixed ISM to be largely true. 

In general, the tail stars exhibit the following trend -- the farther they are from the galaxy disc, the faster and the more metal-poor they are. Their distribution along the cylindrical radius indicates outside-in stripping. Not all stars, however, follow this simple picture, most notably the stars formed from the gas falling back on to the galaxy. Moreover, the star-forming gas can also end up in the galaxy shadow -- a region within the stripping radius where the gas stays protected from the ICM wind -- either by moving there radially or through direct stripping of the central part of the galaxy during the initial phase of the stripping.

Indeed, one of the most interesting insights gained from this figure is the relatively high fraction of stars formed from gas falling back towards the galaxy. Such stars constitute 23 and 15 per cent of all stars formed in the tails of W0 and W45, respectively. This highlights the role of the galaxy shadow, which promotes fallback on to the disc. It may be that in the shadow, the protected gas has better chances to survive and form stars compared to the gas that is constantly exposed to ICM. Since the gas is protected from the said ICM and does not mix with it (which imparts momentum, driving clumps away from the galaxy), the gas will first slow down and then start falling back on the disc, forming stars in its wake. In Sect. \ref{sec5:discussion}, we will go into more detail about the `deviant' stars and what their existence could imply for the observational interpretation of RPS galaxies.

\section{Mock UV images} \label{sec5:uv_maps}

\begin{figure*}
\centering
\includegraphics[width=0.8\textwidth]{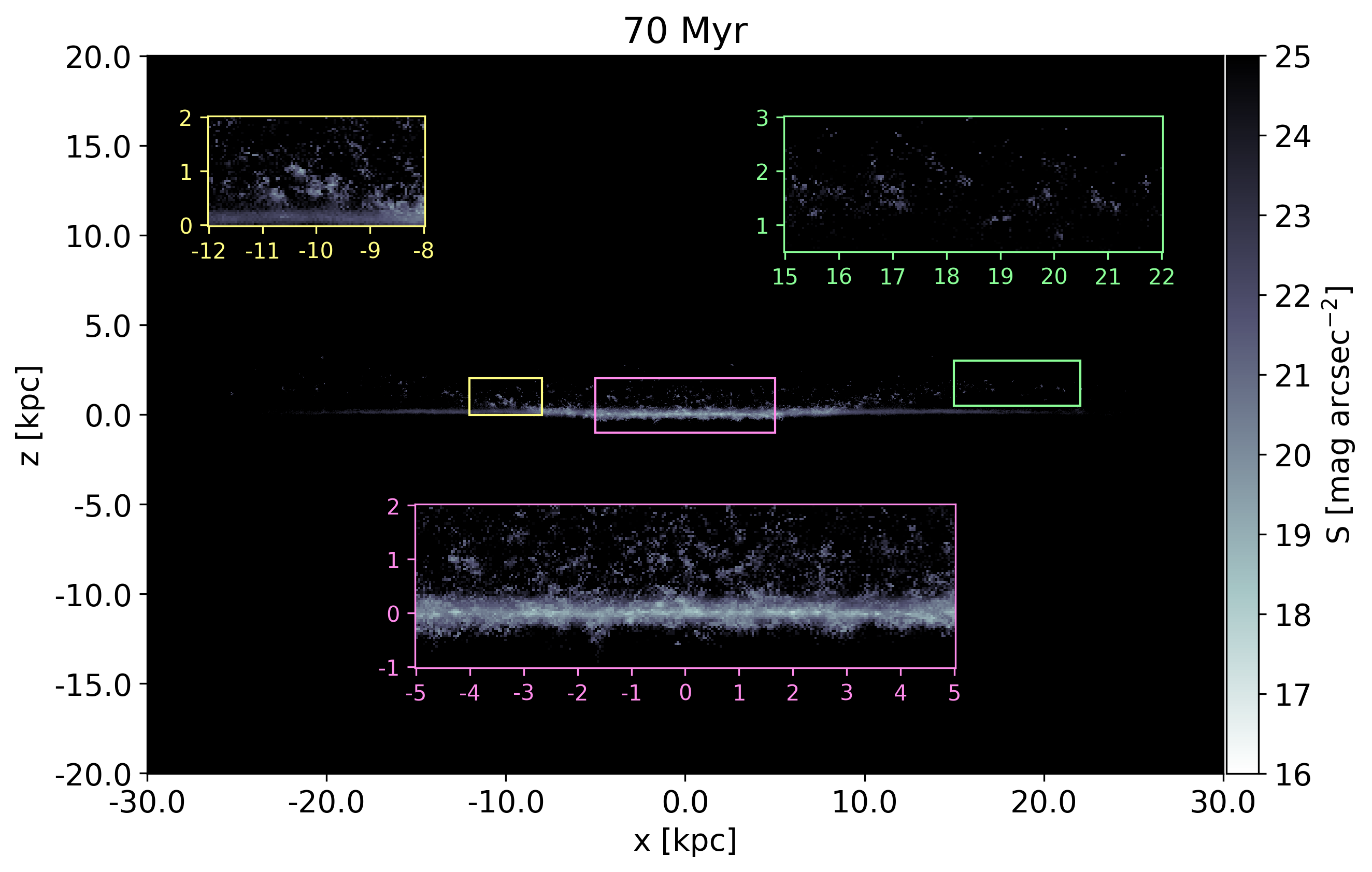}
\caption{Mock UV (HST F275W filter) surface brightness map for W0 at 70 Myr (projection). The upper limit of the surface brightness (colour bar) is the observed HST $5\sigma$ magnitude limit fainter than which we cannot detect the light \citep{Gullieuszik23}. Embedded are three zoomed-in (by 3 times) regions on the map. In this snapshot, the initial ram pressure wave has just passed the galaxy and rapidly stripped a significant portion of gas, which then immediately started to form stars. The pre-RPS stellar population is also visible as a smooth thin disc.}
\label{fig:surface_brightness}
\end{figure*}

Fig. \ref{fig:tail_evolve} reveals that the overwhelming majority of tail stars formed within 20 kpc from the plane of the galaxy end up falling on to the galaxy and seem to form a thick component of the stellar disc. To understand whether this component can actually be observed with current facilities, we have created images that mock UV observations. Specifically, in line with the recent HST observations by \cite{Gullieuszik23} and \cite{Giunchi23a}, we adopt the resolution of UVIS channel of WFC3 camera on board of the HST using the F275W filter. The UVIS pixel scale is of $0.''04$ which corresponds to $dl = 40$ pc at the redshift of JO201's parent cluster Abell 85 \citep[$z = 0.056$,][]{Moretti17}. The point-spread function (PSF) of UVIS has an FWHM of $0.''07$ or $dl_\text{FWHM} = 70$ pc in our case.

Note that the mock images will only include the light of stars formed in the simulations, including the ones formed before RPS began. This means that at 0 Myr (after the onset of RPS) the oldest disc stars are 300 Myr old, and at the end of the simulation their age is 1 Gyr. We do not account for the light of the old stellar disc that is represented in the simulation set-up with a static gravitational potential since the UV flux from a stellar populations older than $\sim 300-400$ Myr provides little contribution the overall UV map \citep{Hao11}. We do not include the gas emission or gas absorption either, as their contribution to the broad-band UV emission is negligible. Finally, the images do not mock instrumental noise and do not account for the dust extinction as we have neither a good subgrid model for dust in the simulations nor dust extinction rates measured with HST resolution. Meanwhile, MUSE (Multi-Unit Spectroscopic Explorer) observations reveal a complex picture of clumpy dust distribution both in the disc and in the tail \citep{Poggianti17_GASP_I}, which would be challenging to reproduce with simple post-processing. We once again emphasise that the primary objective of these images is to compare specifically the stellar distribution in galaxies that did and did not undergo RPS.

Making mock observations is a multi-step process. Firstly, we perform synthetic photometry on Single Stellar Populations (SSPs) from the 2016 update of \cite{BruzualCharlot03} to obtain a table of UV fluxes for SSPs of different ages and metallicities. For each star particle formed, whether in the tail or in the disc, we assign the UV flux as a property based on that star's metallicity and age. Younger and more metal-poor stellar populations emit more UV at fixed stellar mass. Since this flux is for an SSP of mass $1 M_\odot$, we further multiply it by the particle mass. This way, we get the UV flux that every single stellar particle formed in the simulation would emit if it were a real stellar population.

To produce a mock image, we define the image plane as $x = [-30, 30]$ kpc and $z = [-20, 20]$ kpc. We divide it into $n_x \times n_z$ pixels, where $n_x = 60/dl$ and $n_z=40/dl$ ($dl = 0.04$ kpc). Along the line-of-sight ($y$-axis) the data span the whole simulation box. To get the UV flux emitted by each pixel, we sum up the fluxes of all stars located there. Next, to imitate a PSF, we convolve the image with a 2d Gaussian kernel with a standard deviation of $\sigma = dl_\text{FWHM} / \sqrt{8 \ln(2)}$.

The final step is to convert the UV flux to surface brightness in order to easily compare with observations. First, the fluxes $f$ (in Jansky) are converted into AB magnitudes:

\begin{equation}
	\text{mag} = -2.5 \log(f/3631)
\end{equation}

Surface brightness is the apparent magnitude per unit area $A = 0.''04$ (which is the UVIS pixel scale in arcsec):

\begin{equation}
	S = \text{mag} + 2.5 \log(A)
\end{equation}

An example of a surface brightness map is shown in Fig. \ref{fig:surface_brightness} for 70 Myr after the onset of RPS. First, note that the upper limit of the surface brightness (colour bar) is the observed $5\sigma$ magnitude detection limit of $25 \text{ mag} \text{ arcsec}^{-2}$ in the HST observations of \cite{Gullieuszik23}. Zoomed-in (by 3 times) boxes are embedded for three regions of the map.

This snapshot illustrates the initial phase of the stripping: the shock wave has just hit the galaxy and stripped a significant portion of gas, which in turn started forming stars. The outside-in nature of the stripping is clearly visible here: as stars get more displaced from the galaxy centre (yellow box) to the outskirts (green box). Stars that seem to be forming in the galaxy centre (pink box) are, in fact, from the galaxy outskirts and are visible due to projection effects (almost no stars form or end up in the inner 5 kpc, see Figs. \ref{fig:r_z_metallicity} and \ref{fig:cyl_r}).

\begin{figure*}
\centering
\includegraphics[width=\textwidth]{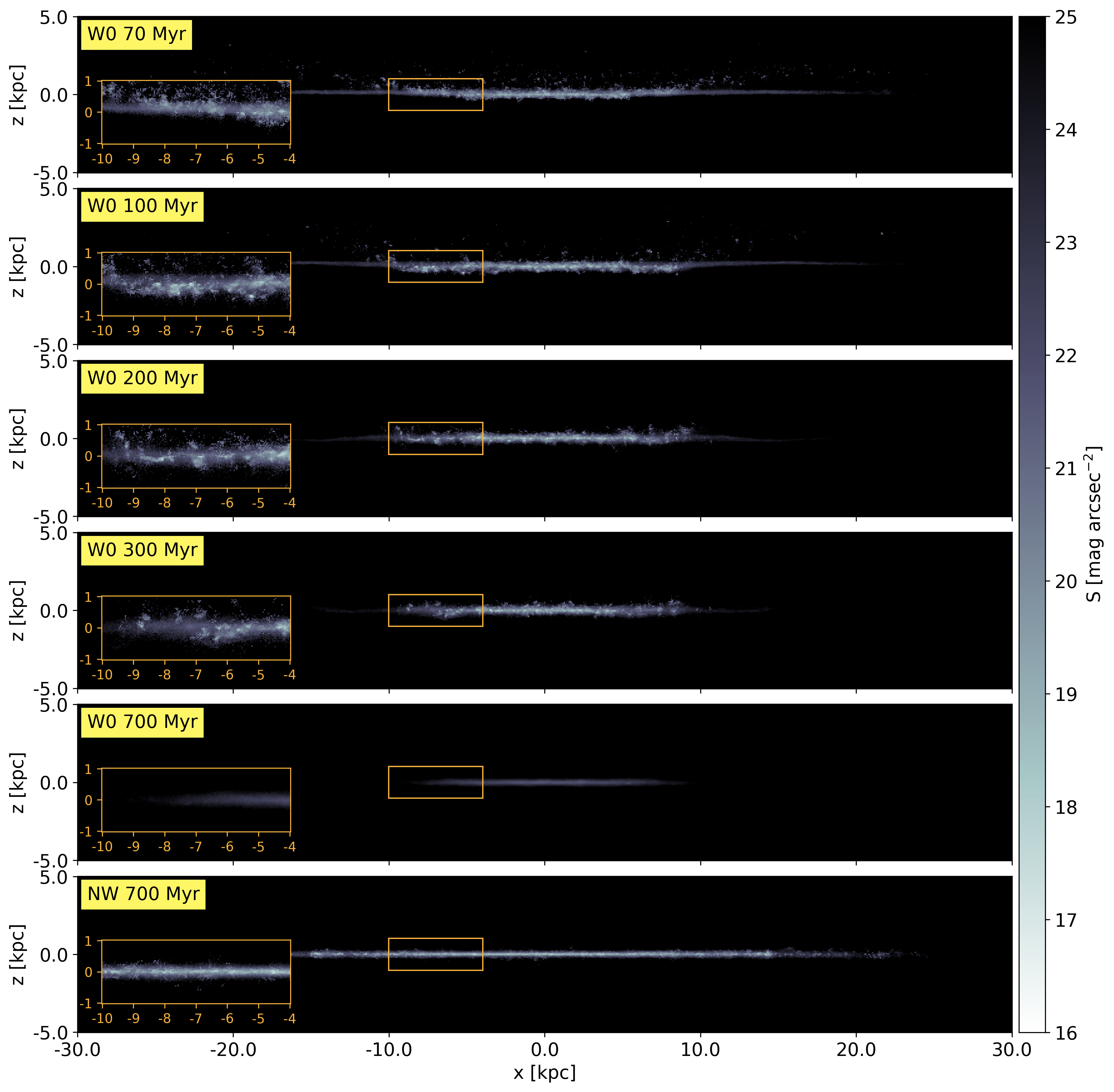}
\caption{Same as Fig. \ref{fig:surface_brightness} but for different galaxies: top four panels are for W0 at 70, 100, 200, 300 and 700 Myr, from top to bottom; bottom panel is for NW galaxy at 700 Myr. Embedded boxes zoom in on the same region in each galaxy. As RPS proceeds, W0 slowly quenches its star formation, and its disc shrinks compared to the NW galaxy.}
\label{fig:surface_brightness_compare}
\end{figure*}

The brightest knots correspond to the youngest stars, and in the galaxy outskirts where the majority of the gas is removed the disc star formation is already quenched. What remains is the dim light of the stars formed prior to RPS. Moreover, while we formally define the galaxy disc with the height of $z=\pm2$ kpc, the real stellar disc is much thinner than that, just about $\pm500$ pc. This is why in this image, although we can hardly find stars much further away than 2 kpc, the tails can still be visually identified. 

Having visually examined all the maps from 0 to 700 Myr with 5 Myr time steps, we can confirm that the mock images do not have tails longer and more pronounced than the ones presented in Fig. \ref{fig:surface_brightness}, even though the real observations show a different picture \citep{Gullieuszik23}. Only in the first $\sim100$ Myr are the stars outside of the galaxy disc clearly visible. Even though, as shown in the previous sections, stars can form at least up to 20 kpc above the disc, they are rare, and the light from a couple of stars formed close together in space is not bright enough to be detected here.

Let us now look at how the UV maps change as the stripping progresses, which we will be comparing to a UV map of an isolated NW galaxy not subject to RPS. In Fig. \ref{fig:surface_brightness_compare} we show in the top four panels maps for W0 at 70, 100, 200, 300 and 700 Myr, from top to bottom; and the bottom panel is for NW galaxy at 700 Myr. Embedded boxes zoom in on the same region in each galaxy. We chose this region to better illustrate the changes close to the disc, while the centre of the galaxy ($-5 \text{ kpc} \leq x \leq 5$ kpc) stays relatively similar (compare the area around $-6 \text{ kpc} \leq x \leq -4$ kpc in the embedded boxes in the top three panels). Hereafter we will focus our analysis on W0, as the equivalent maps for W45 galaxy look nearly identical and do not provide valuable new information. For completeness, we include these maps in Appendix \ref{appendix:uv_maps_w45}.

The 700 Myr map of the NW galaxy is also comprised of all the stars formed during the full 1 Gyr of the simulation time. Unsurprisingly, having visually inspected several of the NW maps at different time steps, we can confirm that this galaxy does not undergo major evolution. Hence, this one map can suffice for the comparison between the W0 images and the NW galaxy at any time.

Overall, the W0 maps illustrate the outside-in quenching of star formation, as the disc slowly shrinks in radial size and becomes progressively dimmer in the outskirts. Tail SFR peaks at $\sim100$ Myr (Fig. \ref{fig:SFH_tail}), so this is the time during which the biggest changes in the light distribution should be visible. While the gas is still actively stripped and the long tails (similar to the 70 Myr map) can be identified, some of the stars have already begun falling on to the galaxy. Having moved `through' the old stellar disc, these stars form a bright structure at $-10 \text{ kpc} \leq x \leq 10$ kpc that is located slightly ($\sim400$ pc) but noticeably below the disc (which is positioned at $z=0$ kpc). Such a peculiar distribution can only be found deep within the truncation radius since in order to be able to fall so quickly after the onset of RPS, these stars must have been born close to the disc. This is a population of stars that formed from gas that was perturbed by ram pressure and gained a vertical oscillation, but was not pushed to large distances above the disc.

The stars will continue oscillating around the galaxy plane and will periodically end up higher (see 200 Myr panel) or lower than the old disc. From 200 Myr onward the SFRs drop down to $\leq0.01 M_\odot \text{ yr}^{-1}$ and the newly-formed stars will rarely appear on the maps. These oscillations around the galaxy centre and the gradual disappearance of the light from the outskirts are going to be the only visible changes in the surface brightness images. Slowly, even these stars will age and stop emitting UV, leaving behind a smaller disc of $R=8$ kpc.

The reason behind making these mock UV maps was to answer the question: can the apparent thickening of the galaxy disc due to the fallback of tail stars be visible in the UV? Looking at Fig. \ref{fig:surface_brightness_compare} it becomes clear that the answer depends on the stripping stage and the availability of a sufficient number of young stars. In any case the difference between stripped and non-stripped discs is minor, and can only be properly seen when zooming-in on the galaxy (as in the embedded boxes). Furthermore, one needs to keep in mind that these maps represent a perfect scenario in which a very thin galaxy disc is viewed exactly edge-on. In any other configuration or in a disc with a larger scale height, this thickening might not be detectable at all, which further decreases the probability of this effect being observed.

\section{Discussion} \label{sec5:discussion}

\subsection{Caveats}

While we analyse the star formation in the tails and the fate of the stars, it is important to keep in mind one caveat in the adopted star formation model. That is, the minimum mass of a stellar particle is $1000 M_\odot$. This choice has several implications. First of all, the actual number of stars formed in the tails is underestimated. This is by design as tracking stars is extremely computationally expensive, but this also means that the exact quantity of stars formed in the tails is not a number to be used as a reference point. Furthermore, the distribution of stars as they fall on to the galaxy could differ depending on the number of stars. For example, compared to the presented map of $\sim6\times10^3$ stars (Fig. \ref{fig:surface_brightness}), $6\times10^4$ stars could form a thicker disc due to increased random motions. The exact outcome is hard to predict, since less massive stars would emit less UV flux and would be distributed differently (i.e. number of star particles per pixel would differ).

A lower minimum mass would lead to a bigger number of stellar particles for two reasons. The first one is obvious. Where before only one stellar particle of mass $1000 M_\odot$ was formed, now ten stars of mass $100 M_\odot$ would appear. The second reason has to do with gas clouds of lower masses. As the gas gets stripped and destroyed, the number of clouds of masses sufficiently high to form massive stellar particles is continuously decreasing. The lower threshold mass could result in higher SFRs, particularly at the end of the simulation when in W0 star formation is totally quenched. A similar argument could be made about the choice of the minimum threshold density of $n_{\rm{thresh}}=10 \, \rm{cm}^{-3}$. While we had previously tested different values and found that they lead to similar galaxy evolution on the global scale \citep{Akerman24}, a lower threshold value could allow for an increased formation of stellar particles in the tails in particular.

Still, we expect statistical results as in Sect. \ref{sec5:SFH_fate} and \ref{sec5:star_formation_how} to be qualitatively independent of the minimum mass of a star particle, and the total stellar mass should not be considerably underestimated either.

All these caveats could have contributed to the differences between the observed \citep{Gullieuszik23, Giunchi23b} and the mock UV images. Notably, the observed UV clumps seem to be bigger ($\sim10^2$ pc) and located at farther distances (15--25 kpc) from their host galaxies. Increased number of formed stars as a result of changes in the star formation model could result in stars forming clusters or clumps similar to the observed ones. Several stars located close together may altogether have higher surface brightness than if only one massive bright star was emitting per pixel (as it happens here). Additionally, the strong ram pressure could also contribute to the distribution of tail stars: the chosen ram pressure might have been overestimated, particularly in the initial values, resulting in the immediate displacement of a considerable amount of gas \citep{Akerman24}. This gas was not stripped far enough and as a result formed stars close to the disc. Had the stripping proceeded more smoothly throughout the simulation, perhaps, the tail stars would also be formed father from the galaxy. The goal of these simulated images, however, was the study of fallback stars and the formation of a thick stellar disc. As we have written above, with so many changes it would be challenging to predict the exact result, but low-mass, old stars (as it takes time for the stars to fall on the galaxy) that do not strongly emit in the UV are unlikely to produce a noticeably thicker distribution, and therefore we expect that more stars forming at large distances and falling back on to the disc would not result in a thick UV-emitting disc.

\subsection{On the radial movement of tail stars} \label{subsec:radial}

\begin{figure}
\centering
\includegraphics[width=\hsize]{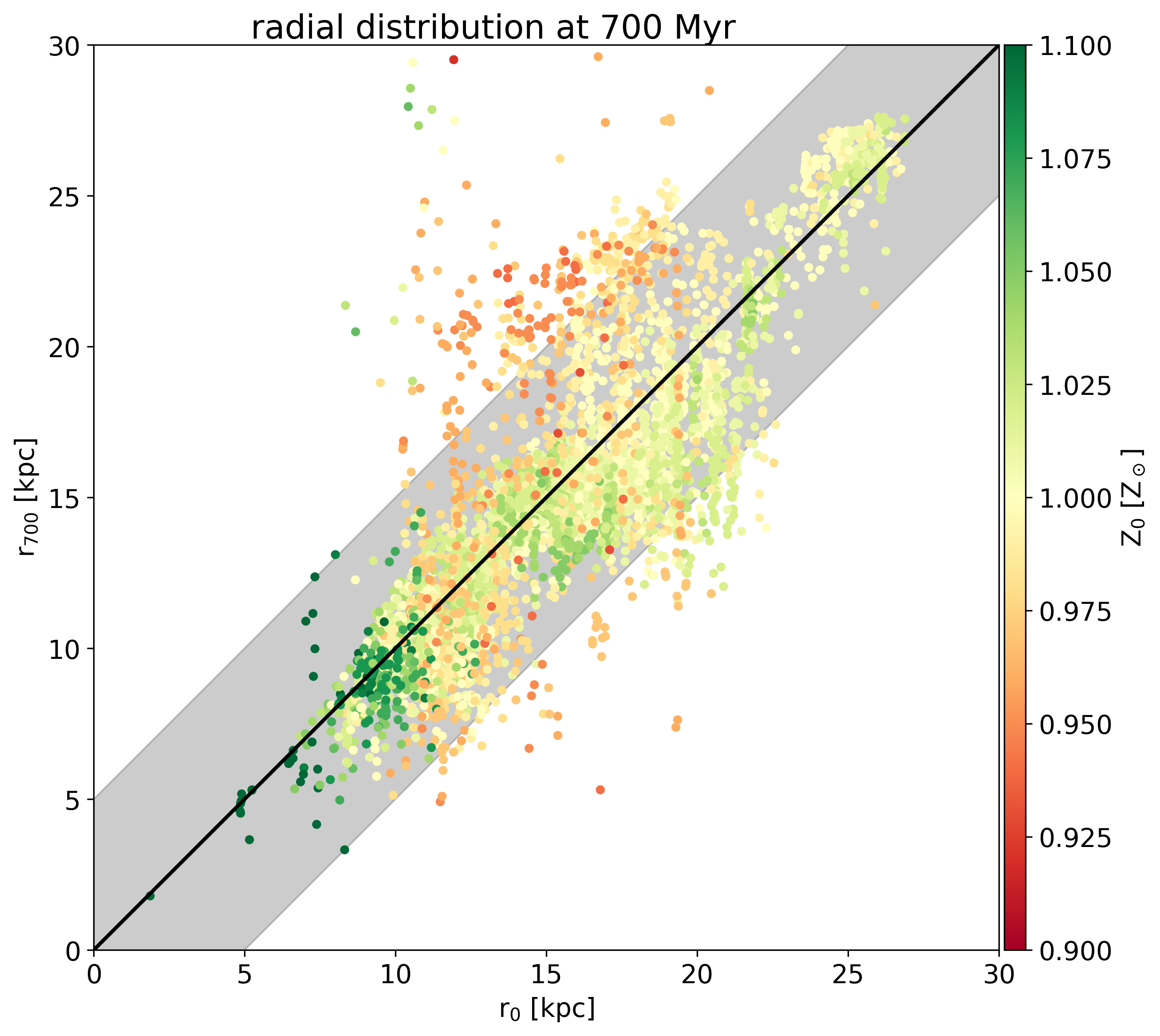}
\caption{For W0: cylindrical $r$ of tail stellar particles at 700 Myr ($y$-axis), where on the $x$-axis is the cylindrical $r$ with which the particles were formed. The points are colour-coded by their metallicity. The black line is the line of equality and the grey area denotes $\pm5$ kpc from this line.}
\label{fig:cyl_r}
\end{figure}

Here, we would like to go back to the problem of (cylindrical) radial evolution of the tail stars that we touched upon in Sect. \ref{sec5:SFH_fate}. Although the upper panel of Fig. \ref{fig:tail_evolve} suggests that after their formation the tail stellar particles simply fall `straight down' on the galaxy, in the same projection we also see red points distributed all the way to small $y$ values -- a clear projection effect. To check whether stars shift in radial coordinates after being formed a new figure is required.

\begin{figure*}
\centering
\includegraphics[width=0.8\textwidth]{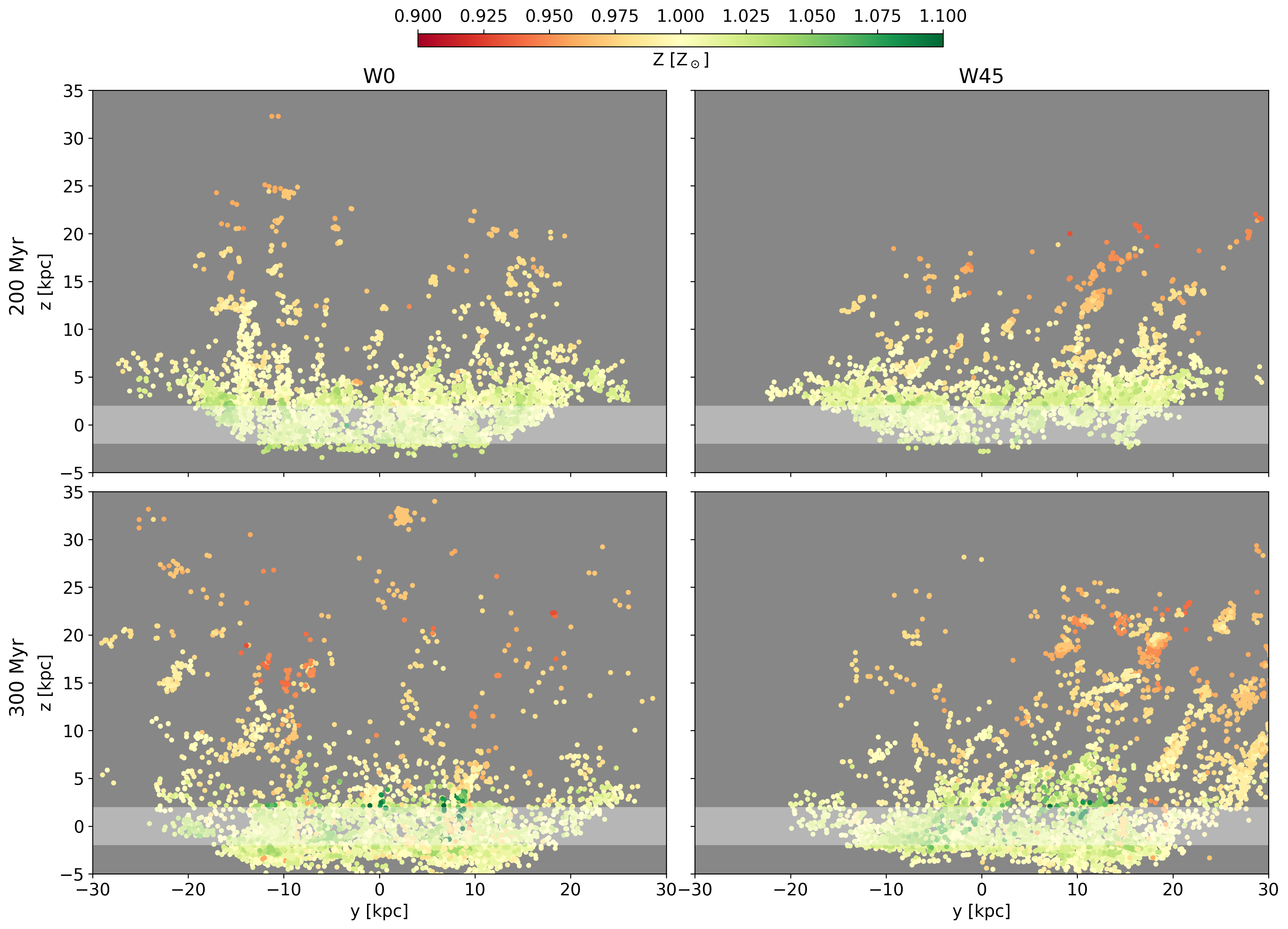}
\caption{The positions of stars at 200 Myr (top) and 300 Myr (bottom) for W0 (left column) and W45 (right column), colour-coded by the metallicity. The white area outlines the disc ($R=30$ kpc, $z=\pm2$ kpc). The picture of tail star formation can be confusing even the case of face-on stripping, where due to projection effects stars on the outskirts of the galaxy appear to be located in its centre. The clean picture of the metallicity gradients, where stars located farther from the disc have lower metallicity due to the prior ISM--ICM mixing, is complicated by the fact that even at the same height stars can be moving in different directions, with some falling back on the galaxy. At later times, this leads to oscillations of stars around the galaxy disc.}
\label{fig:tail_snapshot}
\end{figure*}

In Fig. \ref{fig:cyl_r}, for W0 we compare the cylindrical $r$ coordinates of tail stars at their creation time ($r_0$) with that at 700 Myr ($r_{700}$). We colour the points with the metallicity of the stars to illustrate what gas they were formed from: metallicity above $1.0 Z_\odot$ suggests that the gas was almost `pure' ISM, while values below $1.0 Z_\odot$ indicate some degree of mixing with the ICM. Before we consider the radial motion of star particles, we highlight that at 700 Myr only three star particles are within a 4 kpc (cylindrical) radius of the disc centre, so basically all star particles that seem to be at the disc centre in Figs. \ref{fig:tail_birth} and \ref{fig:tail_evolve} are only there due to projection effects.

From Fig. \ref{fig:cyl_r} we see that the dynamics of tail-born stars is mostly determined by the degree of mixing between the stripped ISM and the low-metallicity ICM. The majority of star particles (80 per cent) end up within 5 kpc of their radius of formation (grey area), which is especially true for the ones formed from `pure' or relatively unmixed gas (green and yellow points). This is because due to the low degree of mixing, the fast-moving ICM has a relatively small effect (in terms of kinematics) on the gas after it was stripped. In fact, this gas stays fairly close to the galaxy: 97 per cent of stars with metallicities $Z_0 \geq 1 Z_\odot$ are born close to the disc ($|z_0|<5$ kpc).

On the contrary, stars that do not simply fall back to their $r_0$ are the ones formed from mixed gas (red and orange points). They are born higher from the galaxy (see Fig. \ref{fig:r_z_metallicity}) and their dynamics is characterised by oscillations around the disc where they move not only along the $z$-axis, but also along the cylindrical $r$. Such changes in the stellar orbits are driven by the ICM imparting momentum on to the parent gas and the changes in the centripetal force which increase and decrease based on the star's distance to the galaxy centre.

\subsection{The picture of star formation in the tails}

The picture of star formation in the tails generally follows from the simple outside-in stripping model. As the gas is being stripped farther and farther away from the galaxy, it gradually mixes with the metal-poor and high-velocity ICM. The result is a gradient of increasing velocities and decreasing metallicities of stars as a function of distance from the galaxy.

In the W0 simulation, the majority of the tail gas is stripped right away and continues forming stars as it is accelerated away from the galaxy, creating a clean picture of metallicity and velocity gradients (Fig.\ref{fig:tail_snapshot}, left column). While the overall picture of gas clouds being stripped, collapsing and forming stars along the way is seen in our simulations, stars are formed at a range of heights above the disc, and even at the same formation height they can have a wide range of velocities. This is because many clouds of different densities and surface areas are being stripped at any given moment, and different cloud properties will result in different acceleration and cooling/star-formation times. This picture is further complicated in W45 galaxy (Fig.\ref{fig:tail_snapshot}, right column), where star formation lasts for a longer period of time, making it difficult to discern exactly when the parent gas cloud of each individual star particle was stripped. Furthermore, the angled wind alters the dynamics of the stripped gas, elongating the orbits and introducing a new radial axis for the gas (and the stars) to travel along. However, even in these snapshots one can see the general trend of decreasing metallicity with increasing distance along the stripped tail.

The motions of the formed stars are not simply dictated by the acceleration of their birth cloud by the ICM wind. The clouds were originally orbiting within the disc, and continue to orbit once they are stripped (Fig. \ref{fig:tail_birth}), creating stellar age gradients that do not necessarily follow the ICM wind direction. In addition, some stars create (apparent) chains that indicate that the parent stripped gas was moving radially outward because of the decrease in the centripetal force as they moved from the disc. Even more dramatically, some stars can be born with negative velocities, indicating that their birth cloud was already falling back towards the galaxy when they were formed -- either because the cloud moved into the shadow of the surviving disc and was protected by the wind or because the gas cloud condensed to the point that the ICM wind could no longer impart enough momentum to continue accelerating it away from the galaxy. All of these motions in directions other than the ICM wind will dilute the predicted velocity and metallicity trends.

Finally, even an observed metallicity gradient with height above the disc can be difficult to interpret. This is because while lower metallicity does indicate mixing between the ISM and metal-poor ICM, in practice, the exact amount of mixed-in ICM in any star-forming cloud is tricky to quantify as there is no reliable way of knowing what the gas metallicity was at the time of stripping. When comparing disc and tail stellar metallicities, the difference between the two (even at the same radius) might be an indication of the metal enrichment of the galaxy disc itself.

Therefore, while a close examination of stellar chains does indicate the trends predicted by simple models, the stars formed throughout a stripped tail tell a much more complicated story of the impact of the initial orbits of dense clouds, hydrodynamical motions affecting the radial position of star-forming clouds, and how fallback can remove most traces of a global velocity- or metallicity-relation with height.

\subsection{Can tail stars contribute to intracluster light?} \label{subsec:ICL}

Intracluster light (ICL) is a diffuse stellar component of a cluster that is gravitationally bound to the cluster as a whole, not the individual galaxies. ICL is extremely difficult to study due to its low surface brightness and challenges in disentangling ICL from the galaxy light (both satellite galaxies and the brightest cluster galaxy). Studies suggest that it constitutes about 5--50 per cent of the cluster stellar mass \citep[depending on the cluster mass,][]{Feldmeier04, Gonzalez07, KrickBernstein07, Castro-Rodriguez09, Kluge21}. \cite{Sun10} argue that RPS could contribute to the ICL if star formation in the stripped tails would proceed with a high enough efficiency. On the other hand, based on observed estimates of SFR in the stripped tails of a large number of galaxies, \cite{Gullieuszik20_GASP_XXI} conclude that the contribution of the stars formed in the stripped tails to the ICL is negligible. Here, we explore this possibility in our simulations. In a previous section we showed that tail star particles eventually fall on to the galaxy disc in our simulations. Real stars, however, could be tidally stripped by the cluster potential. Here we discuss that possibility.

\begin{figure}
\centering
\includegraphics[width=\hsize]{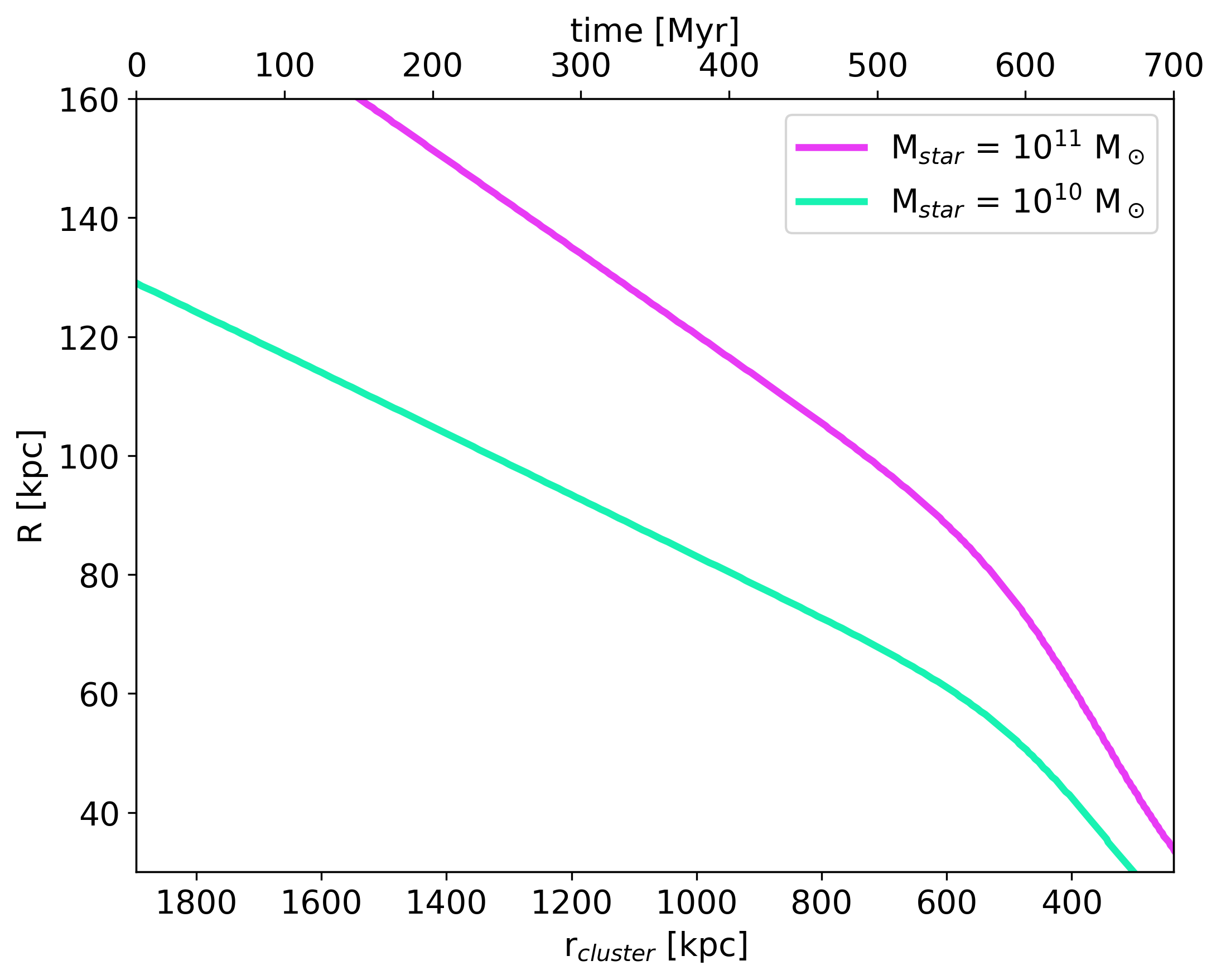}
\caption{Tidal radii of two galaxies as a function of distance to the cluster centre and (the upper $x$-axis) the simulated time of RPS. Purple line represents massive JO201-like galaxy (present in the rest of the paper), while green line is for a smaller, LMC-like galaxy.}
\label{fig:tidal_radius}
\end{figure}

To understand whether in the presence of an additional gravitational potential stars would become unbound from the satellite galaxy we compute the tidal radius shown in Fig. \ref{fig:tidal_radius} for two galaxies. Inside this radius the galaxy's gravity dominates, and cluster's potential prevails outside of it. The tidal radius gradually decreases with time (and clustercentric distance) until about 580 Myr when the galaxy starts to rapidly approach the core of the cluster, where its gravitational pull is much stronger.

Let us first focus on the more massive galaxy (purple line), which is the JO201-like galaxy that we have been working with in the paper so far. We can now count the number of stars that over the course of the simulations were able to get far away enough from their galaxies to be considered tidally stripped. In W0 there are only 39 such star particles (0.66 per cent of the total number), while in W45 the picture is more complex and includes 823 escaped stars (8.9 per cent of the total number). Still, with the average mass of a stellar particle of 1000 $M_\odot$, the stars would only amount to $\sim8\times10^5 M_\odot$. If we consider that a face-on-stripped galaxy loses very few stars, while an edge-on-stripped one has very low stripping rates \citep{Akerman24} and, as a consequence, low tail SFRs, we argue that among all the possible wind angles, W45 should lose the most stars to tidal stripping. With this, we also posit that an average mass of lost stars per galaxy is $\sim5\times10^5 M_\odot$.

\begin{figure*}
\centering
\includegraphics[width=\textwidth]{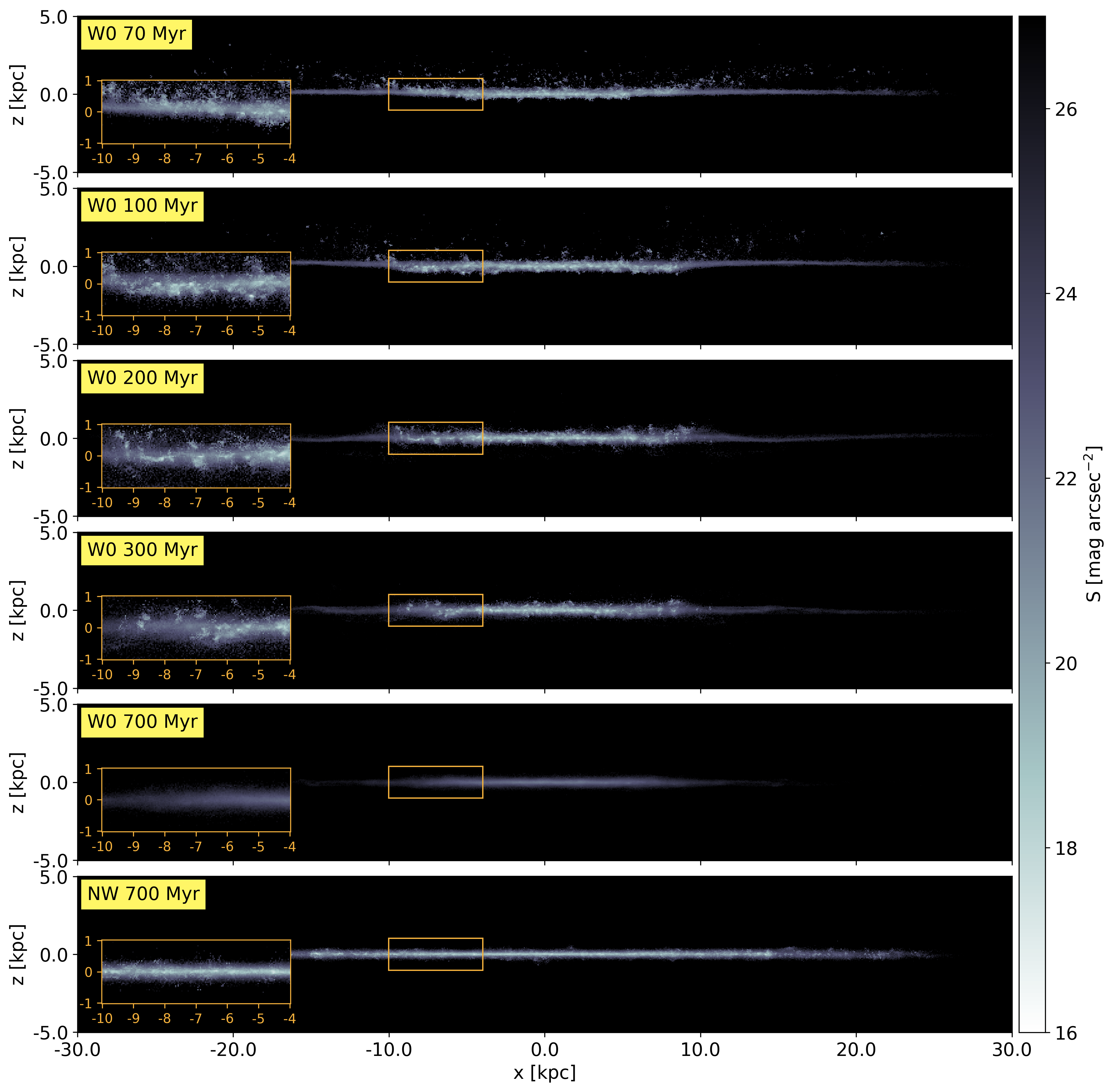}
\caption{Same as in Fig. \ref{fig:surface_brightness_compare} but for surface brightness (colour bar) the upper limit is increased to $27 \text{mag} \text{ arcsec}^{-2}$.}
\label{fig:surface_brightness_new_lim}
\end{figure*}

Let us again consider that ICL accounts for 5--50 per cent of the cluster stellar mass. Depending on the cluster mass, the ICL would amount to $10^{11}$--$10^{13}M_\odot$ \citep[also comparable with results of cosmological simulations][]{Pillepich18, ContiniGu20}. Cluster growth is limited, with clusters usually accumulating double their mass from redshift $z=1$ \citep{vanderBoschFrank02}. Consider two clusters, $10^{14}M_\odot$ and $10^{15}M_\odot$. In order to double their mass, the clusters would have to accumulate 50 and 500 galaxies of mass $10^{12}M_\odot$\footnote{$M_{200}$ of the DM halo of our galaxy is $1.15 \times 10^{12} M_\odot$.}, respectively. This is an extreme upper-limit as mass would likely be accreted both smoothly and from lower mass galaxies as well. Here we select a massive galaxy similar to the one modelled in our simulations, and its mass includes the DM halo. The tidally stripped tail stars from these galaxies would collectively be able to contribute to only 0.025 and 0.002 per cent of ICL in these clusters, respectively.

We can repeat this exercise for a smaller galaxy, although for simplicity we will assume that it has the exact same tail SFR as the more massive JO201. We compute the tidal radius (green line) for an LMC-like galaxy (Large Magellanic Cloud) which similar to the original JO201-like galaxy has the following parameters (see Sect. \ref{sec5:suite}): $M_\text{star} = 10^{10} M_\odot$, $r_\text{star} = 2.7$ kpc, $z_\text{star} = 0.19$ kpc and $r_\text{DM} = 10.89$ kpc. Following our previous calculations, we find that this time, 57 star particles escape in W0 and 998 in W45, and in this case the average mass of lost stars per galaxy is also $\sim5\times10^5 M_\odot$. The two clusters, $10^{14}M_\odot$ and $10^{15}M_\odot$, are able to accumulate 500 and 5000 galaxies, respectively, and the stripped tail stars could contribute to 0.25 and 0.025 per cent of the ICL in these clusters. In this case, the values present an upper limit to the contribution of RPS to the ICL, since a real galaxy of stellar mass $10^{10} M_\odot$ would not have the same tail SFR as a much more massive galaxy.

\cite{Gullieuszik20_GASP_XXI} make an estimation of the contribution of tail star formation to the ICL and conclude that the stars would amount to $4\times10^9 M_\odot$ per cluster since $z=1$. This value was extrapolated from the average tail SFRs and the number of galaxies infalling into a cluster from redshift $z=1$ to the present day. Note that, firstly, the observed SFRs are an order of magnitude higher than the ones presented here and, secondly, while \cite{Gullieuszik20_GASP_XXI} count all of the the tail star formation, here we only consider those stars that could be tidally stripped from a galaxy, and so their number is also an upper limit of how much RPS contributes to ICL.

While we cut the tail region at $z=20$ kpc, we expect that stars would otherwise form at much higher $z$ as in \cite{Kapferer09} and \cite{TonnesenBryan12}. We note that in \cite{TonnesenBryan12}, the tail SFR from 2 kpc $<$ z $<$ 20 kpc is similar to that found in this work, and the total SFR above 20 kpc is less than that in the lower tail. If that trend were to hold here, the stellar mass in the distant tail would therefore be low. Indeed, even if stars were forming as far away as 120 kpc, they would still be mainly gravitationally attracted to the galaxy until 400 Myr (Fig. \ref{fig:tidal_radius}), at which point the SFR in the tails almost completely quenches.

Thus we conclude that the tail stars do not majorly contribute to the ICL (at least for a massive galaxy with a strong gravitational field such as JO201). 

\subsection{Observations of tail stars}

We have shown that the current observational capabilities of the WFC3 camera on board of the HST are not enough for the detection of disc thickening as a result of RPS, at least in the UV band. Would an eventual successor to WFC3/HST capable of reaching a deeper surface brightness be able to detect this effect? To answer this question, let us assume that the spatial resolution remains unchanged. This hypothetical camera cannot be unrealistically better than current-generation instruments, but an increase of two orders of magnitude (from 25 to $27 \text{ mag} \text{ arcsec}^{-2}$) is within technological capabilities.

In Fig. \ref{fig:surface_brightness_new_lim} we present the same maps as in Fig. \ref{fig:surface_brightness_compare} only with the increased surface brightness (colour bar) limit. These two figures look surprisingly similar when looking at the vertical distribution of stars, although the enhanced overall brightness and the radial extension of the disc can be noticed. Only in the 100 and 200 Myr panels, the W0 disc is visibly different from the NW one, being twice as thick. Unfortunately, the thickening of up to 1 kpc is not nearly close to the expected 5 kpc from Fig. \ref{fig:tail_evolve}. More importantly, this $\sim100$ Myr time window is very short compared to the whole duration of stripping. Paired with the fact that the thickening (small as it is) can only be visible in close to edge-on disc configurations, this renders an actual observation improbable. This conclusion is valid only for the UV observations.

Although the effect of stars falling on the galaxy cannot be observed in the edge-on configuration, it could, perhaps, be visible face-on. Chaotic movements of stars as they oscillate around the galaxy plane might lead to an increased velocity dispersion, especially in the galaxy centre, compared to a galaxy in isolation. While quantifying or modelling this effect is out of scope of this work, this is a possible avenue for future studies.

Fig. \ref{fig:tail_evolve} reveals another potentially observable effect, this time about the W45 galaxy. Since in this case the stars do not simply fall straight down on the galaxy disc, but go on parabolic orbits due to the additional $v_y$ component, at the end of the simulation, these stars can gather under the disc. They can be easily identified in Fig. \ref{fig:tail_evolve}, where they are arranged in a structure similar to a tail, but which is mirrored with respect to the direction of ICM wind (which was flowing from the bottom left to the top right corner in this projection). While this effect is unlikely to be observed in the UV, due to the fact that by the time the stars reach the opposite side of the disc they are already too old to be bright in UV, it could potentially be observed at longer wavelengths, e.g. in the B-band.

This stellar excess on the side facing the wind has several implications. Although in this specific image at 700 Myr all the gas has already been stripped, at earlier times the galaxy could appear to have two tails: gas pointing in the direction of stripping and a stellar tail pointing in the opposite direction. Keep in mind, that in this simulation the galaxy is always on its first infall into the cluster, and has not yet passed the pericentre, but can already exhibit asymmetries in gas and stellar distributions. Moreover, when the gas tails are no longer present, such distributions of stars could be mistaken as to having been formed from the stripped gas right in that specific area, and thus the direction of stripping (and the direction in which the galaxy is moving) could be misinterpreted to be the opposite of what it truly is. This tail could also mistaken to be a feature of a tidal interaction. That being said, as with the thickening of the disc in W0, the probability of observing this effect heavily depends on the disc configuration and stripping stage, and requires the galaxy to be positioned nearly edge-on.

While in the two simulations studied here the stellar distribution is not significantly different from an isolated disc in the UV band, we have found interesting possible avenues for follow-up work, such as making mock edge-on H$\alpha$ and B-band maps (similar to the UV maps presented here) and face-on maps of the velocity structure. We also find that the stellar distribution can actually be weighted on the side in the direction of motion depending on the stripping stage.  

\section{Conclusions} \label{sec5:conclusions}

In this paper we analyse star formation in the stripped tails (up to 20 kpc from the galaxy plane) formed behind a massive galaxy ($M_\text{star}=10^{11} M_\odot$). We primarily focus on two RPS configurations: face-on stripping (W0) and angled wind (W45). The main results are the following:

\begin{enumerate}

\item More stars and on a longer time scale are formed when an angled wind is present (58 per cent difference). The excess star formation can be attributed to the combined action of the ram pressure and gas orbit that makes the gas slow down and pile up in one sector of the galaxy (Sect. \ref{sec5:SFH_fate}).

\item Most of the stars are born close to the disc in the first 100--200 Myr as a result of the initial jump in wind strength passing through the galaxy and displacing a lot of gas. Generally speaking, newly-formed stars follow the well-known outside-in stripping scenario (Fig. \ref{fig:tail_birth}). The stripped gas gradually mixes with the ICM, which results in the stars displaying metallicity and velocity gradients along the wind direction (Fig. \ref{fig:r_z_metallicity}). 

\item This simple model, however, does not always hold even in the straightforward face-on stripping case (Sect. \ref{sec5:star_formation_how}). Some gas ends up in the galaxy shadow where it stays protected from the ICM wind, which complicates predicting its further evolution. Part of this protected gas can fall back on the galaxy. Along its way, the gas will continue forming stars (15--25 per cent of the total number of tail stars are born with $v_{z0}<0$). Such stars can be preferentially found in the inner galaxy regions, although they do not usually reach the central few kpc, and close to the disc.

\item Under the influence of gravity, almost all tail stars will end up falling back on to the galaxy (Fig.~\ref{fig:tail_evolve}). Only a few stars are able to reach the escape velocity. Furthermore, having calculated the stripping radius (Sect. \ref{subsec:ICL}), we can confirm that only a small part of tail stars (and only in W45) could be considered tidally stripped by the cluster. We conclude that RPS cannot majorly contribute to the intracluster light.

\item The stars that fall on the galaxy form a thicker disc, particularly in the inner regions of the galaxy. Mock observations that mimic images taken in the F275W filter with UVIS/WFC3 on board of the HST reveal that this thickening is unlikely to be observationally detected in the UV (Fig. \ref{fig:surface_brightness_compare}).

\item In the presence of an angled wind (W45), stars fall on the galaxy following parabolic orbits, and can even form structures similar to tails on the side that faces the incoming wind (Fig.~\ref{fig:tail_evolve}). This can create an effect at which stellar and gas tails will be facing in the opposite directions even when the galaxy has yet to reach the pericentre of its orbit. In the absence of gas tails (at late stages of RPS, when almost no gas remains in the galaxy), we predict that, if observable, these stellar tails could lead to a misinterpretation of the direction of stripping or see that as a feature of a tidal interaction.

\end{enumerate}

In our previous works \citep{Akerman23, Akerman24} we showed that globally, on the galaxy scale, W0 and W45 galaxies evolve remarkably similarly \citep[see also][]{RoedigerBruggen06}. The striking difference between the histories of their tail star formation (Fig. \ref{fig:SFH_tail}) and the dynamics of the formed stars (Fig. \ref{fig:tail_evolve}), suggests that the angle at which the galaxy moves through the ICM impacts the formation, final configuration and fate of tail stars. Real galaxies, however, rotate as they fall into the cluster, and in light of this work, it proves difficult to predict how the stars would behave in such a scenario. Future works should focus on modelling an ICM wind with a varying angle and comparison to constant-angle simulations and observations.

Although in these simulations most tail stars fall back to the disc but are not bright enough to observably thicken the UV disc, we highlight that this is only a single galaxy and a single orbit. Simulation suites that vary the ICM and orbital properties of satellites will be able to put more constraints on the stellar discs and ICL contribution from realistic populations of cluster satellites.

\begin{acknowledgements}
N.A. would like to thank Marco Gullieuszik and Neven Tomičić for the useful comments. This project has received funding from the European Research Council (ERC) under the European Union’s Horizon 2020 research and innovation programme (grant agreement No. 833824). The simulations were performed on the Frontera supercomputer operated by the Texas Advanced Computing Center (TACC) with LRAC allocation AST20007 and on the Stampede2 supercomputer operated by the Texas Advanced Computing Center (TACC) through allocation number TG-MCA06N030. B. V. acknowledges the INAF GO grant 2023 `Identifying ram pressure induced unwinding arms in cluster spirals' (PI Vulcani). We use the yt python package \citep{yt} for data analysis and visualisation.
\end{acknowledgements}

\bibliographystyle{aa}
\bibliography{GASP} 

\begin{appendix}
\onecolumn
\section{Metallicity distribution throughout the disc} \label{appendix:metallicity}

\begin{figure}[h!]
\centering
\includegraphics[width=0.6\textwidth]{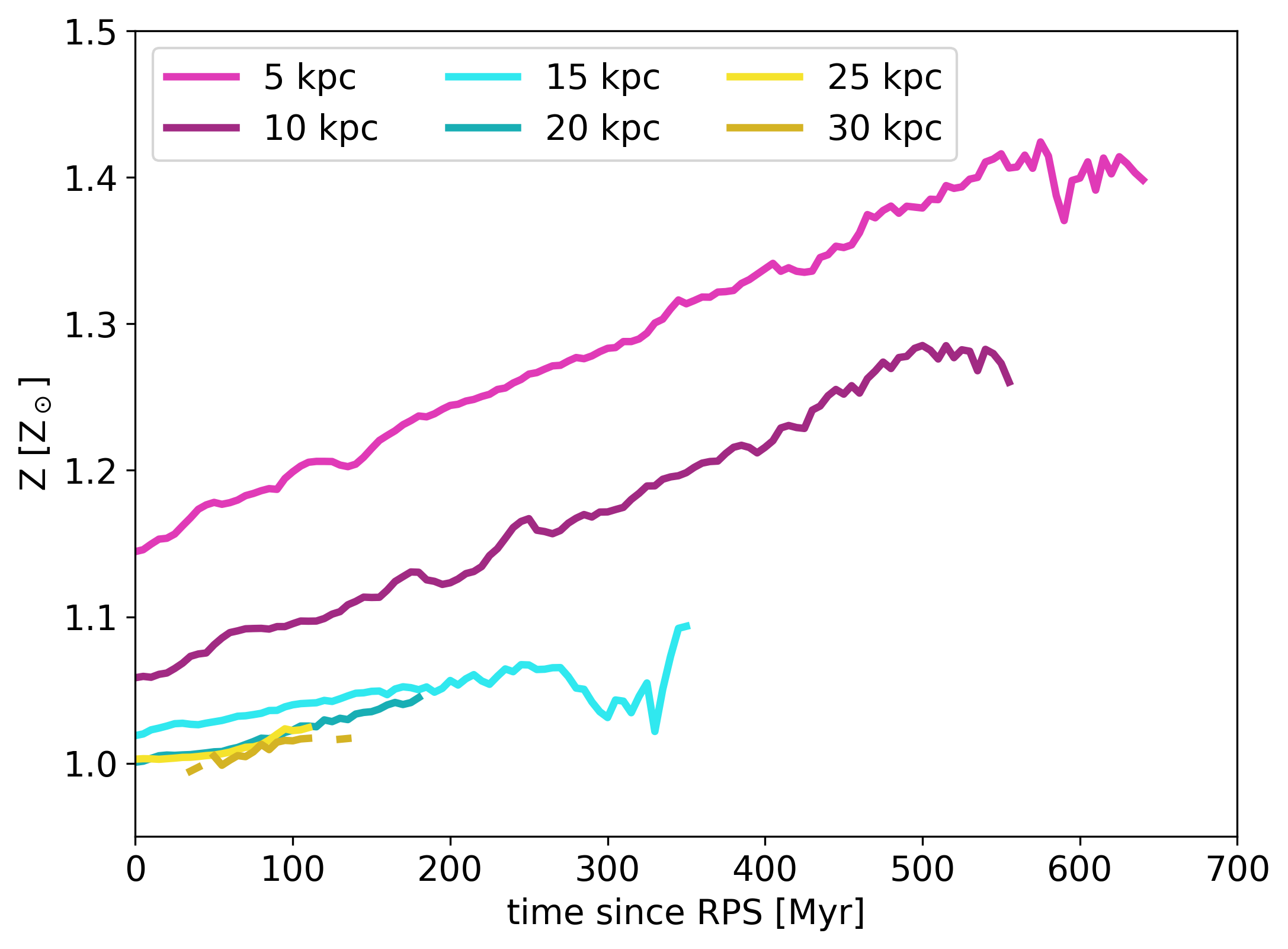}
\caption{Disc metallicity in rings as a function of time for NW galaxy. Metallicity is the average metallicity of young stars ($\leq5$ Myr) at each time step. The lines stop once the gas is stripped from each ring and no new stars can be formed. The $x$-axis is labelled `time since RPS' for ease of comparison with the stripped galaxies discussed in this paper, although no wind was introduced in this case.}
\label{fig:disc_metallicity_rings}
\end{figure}

Here, we show the radial metallicity distribution in the galaxy disc of the NW galaxy. At each time step we find the average metallicity of newborn stars in five concentric rings with the following radii: $R_5 \leq 5$ kpc, $5 \text{ kpc} < R_{10} \leq 10$ kpc, $10 \text{ kpc} < R_{15} \leq 15$ kpc, $15 \text{ kpc} < R_{20} \leq 20$ kpc, $20 \text{ kpc} < R_{25} \leq 25$ kpc and $25 \text{ kpc} < R_{30} \leq 30$ kpc. We refer to these rings by their upper boundary and present the results in Fig. \ref{fig:disc_metallicity_rings}. The height of each ring is the same as the disc definition ($z=\pm2$ kpc). Stars with ages $\leq 5$ Myr are considered newly-born since this is the time step with which the data are saved. This process gives a baseline metallicity as a function of cylindrical radius and time. Note that the $x$-axis is labelled `time since RPS' for ease of comparison with the stripped galaxies discussed in this paper, although no wind was introduced in this case. We can see that the metallicity in the disc is not evenly distributed and the galaxy centre is more metal-rich than the outskirts.

Between the two rings that survive the longest, 5 and 10 kpc, the metallicity difference is consistently $\sim0.1 Z_\odot$, while the metallicity of these two rings increases by $0.03-0.07 Z_\odot$ every 100 Myr. In the 15 kpc ring the slope is smaller and the increase is about $0.02 Z_\odot$ in 100 Myr.

With regards to Sect. \ref{sec5:star_formation_how}, if the $1.5 Z_\odot$ gas gets stripped and then mixes with ICM, the resulting gas can still have metallicity above $1.0 Z_\odot$, and would be considered unmixed in an initial analysis of Fig. \ref{fig:r_z_metallicity}. From Fig. \ref{fig:disc_metallicity_rings} we see that the gas stripped from the outskirts and the gas stripped from the inner disc would have different metallicities even if the stripping occurred at the same time. However, we can also see that in the outer regions of the galaxy ($r>10$ kpc) the metallicity increases very slowly once RPS is introduced and stays around 1.0--1.05$Z_\odot$. Since most of the stars are born outside of this radius, we find our assumption that metallicities $Z>1 Z_\odot$ are an indication of the `pure' or unmixed ISM to be largely true.

\newpage
\section{Mock UV maps for W45 galaxy} \label{appendix:uv_maps_w45}

\begin{figure}[h!]
\centering
\includegraphics[width=\textwidth]{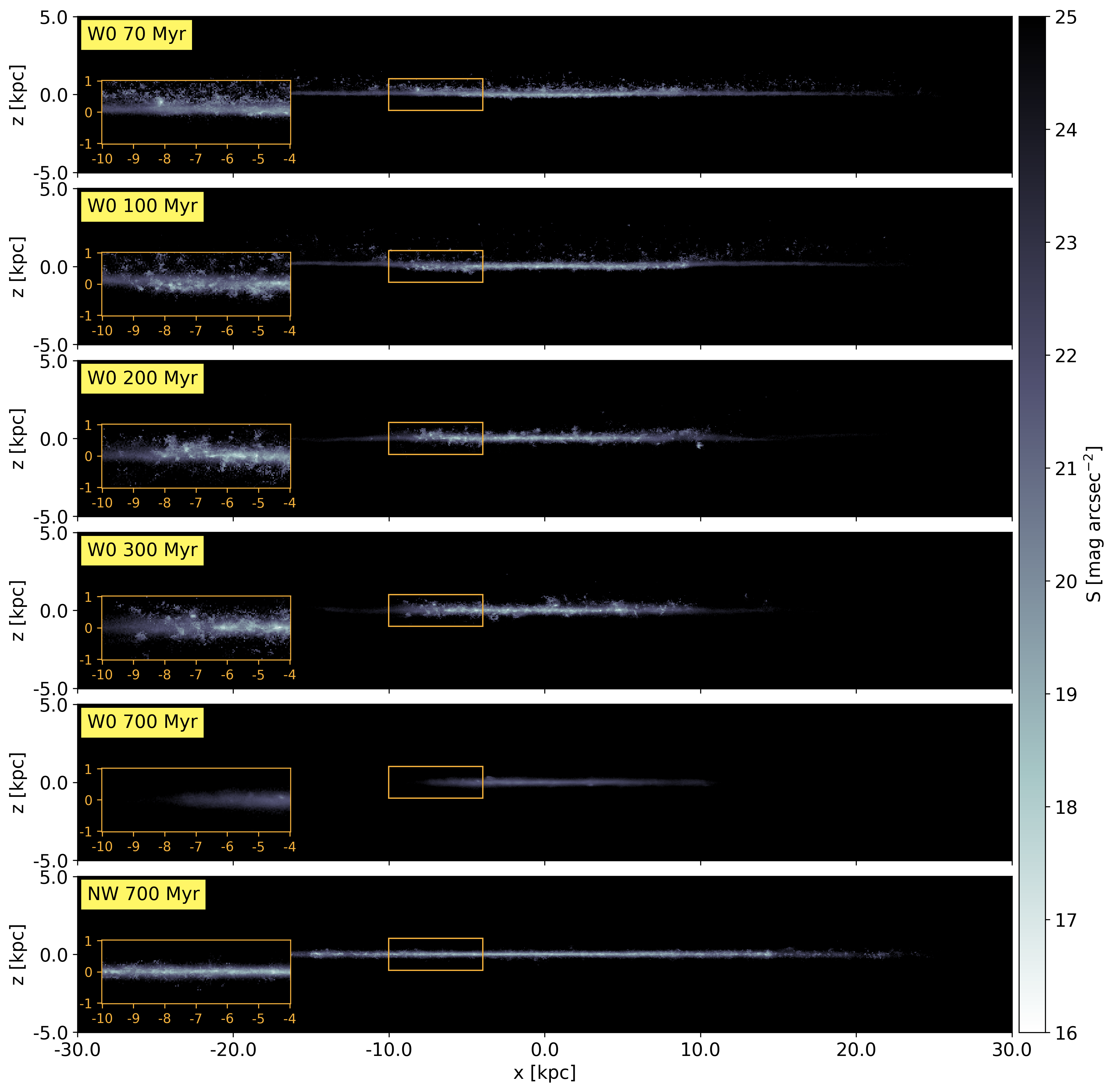}
\caption{Same as Fig. \ref{fig:surface_brightness_compare} but for W45 galaxy.}
\label{fig:w45_surface_brightness_compare}
\end{figure}

Here in Fig. \ref{fig:w45_surface_brightness_compare} we show the mock UV maps for W45 galaxy, which is the same as Fig. \ref{fig:surface_brightness_compare} for W0. Surprisingly, despite the differences in star formation and stripping histories, at the respective times the W0 and W45 maps look remarkably similar, even within the embedded boxes. Even more surprising is how symmetric the W45 looks in all of the frames, with only slight differences between the leading and the trailing edges of the galaxy. This could be explained with the fact that on both edges of the galaxy the stars will start oscillating around the disc (which is also rotating), smoothing out the overall surface brightness. Additionally, just as in W0, the remarkable tails which exemplify the asymmetry in the stripping are not visible in the mock maps, because of the low concentration of tail stars.

\end{appendix}
\end{document}